\documentclass[12pt,epsf,qsymbols]{article}
\usepackage[utf8]{inputenc}
\usepackage[normalem]{ulem}
\usepackage[english]{babel}%,frenchb
\usepackage{tabularx}
\usepackage{array}
\usepackage{graphics}
\usepackage[pdftex]{graphicx}
\usepackage{psfrag}
\usepackage{epsfig}
\usepackage{subfig}
\usepackage{mathtools}
\usepackage{amssymb}

\usepackage{setspace}
\usepackage{rotating}
\usepackage{colortbl}
\usepackage{longtable}
\usepackage{slashed}
\usepackage{braket}
\usepackage{lineno}
\makeatletter
\usepackage{textcomp}
\usepackage[usenames,dvipsnames,table]{xcolor}
\usepackage{relsize}
\usepackage{cite}

\usepackage{float}

\usepackage{bbold}
\usepackage{bbm}
\usepackage{bm}
\usepackage{enumerate}
\usepackage{amsmath}
\usepackage{verbatim}
\usepackage[normalem]{ulem}
\usepackage{hyperref}
\hypersetup{colorlinks,citecolor=nicegreen,linkcolor=niceblue}
\hypersetup{colorlinks=true}

\allowdisplaybreaks

\setlongtables

\newcommand{\cb}{{\mathcal B}}
\newcommand{\bea}{\begin{eqnarray}}
\newcommand{\eea}{\end{eqnarray}}
\newcommand{\beq}{\begin{equation}}
{
\newcommand{\eeq}{\end{equation}}
\newcommand{\ec}{\end{center}}
\newcommand{\bc}{\begin{center}}

\newcommand{\gev}{{\rm GeV}}
\newcommand{\mev}{{\rm MeV}}

\newcommand{\pdir}{p\kern -5.2pt\raise 0.2ex\hbox {/}}

\newcommand{\vdir}{v\kern -5.75pt\raise 0.15ex\hbox {/}}
\newcommand{\kdir}{k\kern -5.75pt\raise 0.15ex\hbox {/}}
\newcommand{\epsdir}{\epsilon\kern -5.0pt\raise 0.15ex\hbox {/}}
\newcommand{\bvdir}{\bar{v}\kern -5.75pt\raise 0.15ex\hbox {/}}
\newcommand{\Ddir}{D\kern -7.75pt\raise 0.20ex\hbox {/}}
\newcommand{\Adir}{A\kern -7.75pt\raise 0.20ex\hbox {/}}
\newcommand{\ldir}{l\kern -5.0pt\raise 0.2ex\hbox{/}}
\newcommand{\varepsdir}{\varepsilon\kern -5.5pt\raise 0.15ex\hbox{/}}

\newcommand{\nn}{\nonumber}
\makeatother

\definecolor{niceblue}{rgb}{0.15,0.15,0.6}
\definecolor{nicegreen}{rgb}{0.1,0.5,0.1}
\definecolor{Red}{rgb}{1.,0.,0.}

\definecolor{Green}{rgb}{0.2,.7,0.2}

\begin{document}
\unitlength = 1mm

\thispagestyle{empty} 
\begin{flushright}
\begin{tabular}{l}
{\tt \footnotesize LPT-Orsay-17-18}\\
\end{tabular}
\end{flushright}
\begin{center}
\vskip 2.4cm\par
{\par\centering \textbf{\LARGE  
\Large \bf Seeking the CP-odd Higgs via $h\to \eta_{c,b}\ell^+\ell^-$ }}
\vskip 1.2cm\par
{\scalebox{.85}{\par\centering \large  
\sc Damir Be\v{c}irevi\'c$^a$, Bla\v{z}enka Meli\'c$^b$, Monalisa Patra$^b$ and Olcyr Sumensari$^{a,c}$}
{\par\centering \vskip 0.7 cm\par}
{\sl 
$^a$~Laboratoire de Physique Th\'eorique (B\^at.~210)\\
CNRS, Univ. Paris-Sud, Universit\'e Paris-Saclay, 91405 Orsay, France.}\\
{\par\centering \vskip 0.25 cm\par}
{\sl 
$^b$~Institut Rudjer Bo\v{s}kovi\'c, Division of Theoretical Physics,\\
Bijeni\v{c}ka 54, HR-10000, Croatia.}\\
{\par\centering \vskip 0.25 cm\par}
{\sl 
$^c$~Instituto de F\'isica, Universidade de S\~ao Paulo, \\
 C.P. 66.318, 05315-970 S\~ao Paulo, Brazil.}\\

{\vskip 1.65cm\par}}
\end{center}

\begin{abstract}
We show that the decay rates of the Higgs boson to a pseudoscalar quarkonium and a pair of leptons, $h\to P\ell^+\ell^-$ ($P\in\{\eta_c,\eta_b\}$), 
can be substantially enhanced in a scenario with two Higgs doublets with a softly broken $\mathbb{Z}_2$ symmetry (2HDM) when the CP-odd Higgs $A$ is light, i.e. $m_A\lesssim m_h$. Depending 
on the type of 2HDM the enhancement of $\cb(h\to \eta_{c,b}\tau^+\tau^-)$ with respect to its Standard Model value can be an order of magnitude larger, i.e. $\mathcal{O}(10^{-6}\div10^{-5})$. The decays $h\to P\ell^+\ell^-$ could therefore provide an efficient channel to investigate the presence of a light CP-odd Higgs $A$ and help to disentangle among various 2HDM scenarios.\\
\end{abstract}
\setcounter{page}{1}
\setcounter{footnote}{0}
\setcounter{equation}{0}
%%%%%%%%%%%%%%%%%%%%%%%%%%%%%%%%%%%%%%%%
\noindent

\renewcommand{\thefootnote}{\arabic{footnote}}
%\linenumbers

\setcounter{footnote}{0}

%\tableofcontents

%\newpage

\vskip 0.85cm

%%%%%%%%%%%
%%%%%%%%%%%
%%%%%%%%%%%
%\section{Introduction}
%\label{sec:intro}

\section{Introduction}

One of the minimal extensions of the Standard Model consists in enlarging the Higgs sector and, instead of one doublet of scalar fields, introducing an additional Higgs doublet, which comes under the generic name of Two Higgs Doublet Model (2HDM)~\cite{Branco:2011iw}. With a peculiar choice of Yukawa couplings, 2HDM is embedded in the minimal supersymmetric extension of the Standard Model (MSSM) which made it particularly popular in phenomenological applications of low energy supersymmetry. The fact that the mass of observed Higgs boson was found to be consistent with the Standard Model expectations made MSSM less compelling but the 2HDM remains a convenient framework to study the extensions of the Higgs sector in view of the current experimental searches. 

Despite its minimalism 2HDM has a rich structure. Besides the usual Higgs state $h$, there is an extra CP-even scalar ($H$) and two additional states of which one is a charged scalar boson ($H^\pm$) and another a CP-odd state ($A$). The state observed at LHC is identified with $h$~\cite{Aad:2015zhl}, or $H$~\cite{Enberg:2016ygw} which is otherwise expected to be heavier than $h$. A fact that the measured rates of the tree level weak decays of leptons and mesons are consistent with the Standard Model predictions is a hint that the charged Higgs is heavy too, also confirmed by the direct searches~\cite{Akeroyd:2016ymd}. The  CP-odd Higgs boson, instead, remains unconstrained explicitly. It is often assumed that its mass is larger than that of the observed Higgs, $m_A >m_h$. This, however, is just an assumption which should be tested experimentally. 
Several proposals to look for a light CP-odd Higgs were made in the past~\cite{Dermisek:2010mg,Ellwanger:2011sk,Domingo:2008rr,Haisch:2016hzu}. In this paper we propose 
to study decays of Higgs to the pseudoscalar heavy quarkonia ($P$), $h\to P\ell^+\ell^-$, ($P\in\{\eta_c,\eta_b\}$) processes in which the CP-odd Higgs can contribute at the tree level and make a significant enhancement of various decay rates. The level of such an enhancement is related to the structure of the Yukawa couplings and to the mass of the $A$-state.
 As we shall see in the following, we find that $\cb (h\to \eta_{c,b}\tau^+\tau^-)$ can be enhanced by an order of magnitude with respect to its Standard Model value, which is why we find it interesting and worth studying in experiments.  

Studies of the Higgs boson to quarkonia attracted quite a bit of attention: Radiative decay $h\to J/\psi \gamma$ could be used to probe the Yukawa coupling $hc\bar c$, a possibility which is compromised in the case of $b$-quark quarkonia ($\Upsilon(nS)$) due to cancellation of two contributions to the decay amplitude~\cite{Bodwin:2013gca}. A possibility to study $h\to PZ$ and $h\to VZ$ (where $V$ and $P$ stand for the vector and pseudoscalar quarkonium states, respectively) was elaborated in Refs.~\cite{Gao:2014xlv,Alte:2016yuw}. Finally, a possibility to search for the effects of lepton flavour violation via $h\to V\ell_1\ell_2$ has been proposed in Ref.~\cite{Colangelo:2016jpi}.

The remainder of this paper is organized as follows: In Sec.~\ref{sec:h2p} we derive the expressions for $\cb (h\to PZ)$ and $\cb (h\to P\ell^+\ell^-)$ both in the Standard Model and in 2HDM with a light CP-odd Higgs state.  In Sec.~\ref{sec:2hdm} we scan the parameter space of 2HDM with a softly broken $\mathbb{Z}_2$ symmetry. 
The results of our scan are used in Sec.~\ref{sec:eff} where we test the sensitivity of $\cb (h\to \eta_{c,b} \ell^+\ell^-)$ on the presence of a light CP-odd Higgs state. We summarize our findings and briefly conclude in Sec.~\ref{sec:concl}.
 
\section{Expressions for $\cb (h\to PZ)$ and $\cb (h\to P\ell^+\ell^-)$}
\label{sec:h2p}
In this Section we provide the expressions for $\cb (h\to PZ)$ and $\cb (h\to P\ell^+\ell^-)$ both in the Standard Model (SM) and in the generic 2HDM scenario. To do so we need to specify the notation. By $P$ we denote the pseudoscalar quarkonium carrying momentum $k$, while $Z$ flies with momentum $p_Z$. The dilepton invariant mass squared in the $P\ell^+\ell^-$ final state is considered as $q^2 = (p_\ell + p_{\bar{\ell}})^2$, with $p_{\ell,\bar{\ell}}$ being the momentum of the outgoing leptons.

\subsection{$\cb (h\to PZ)$}
\begin{figure}[h!]
  \centering
  \includegraphics[width=1.00\textwidth]{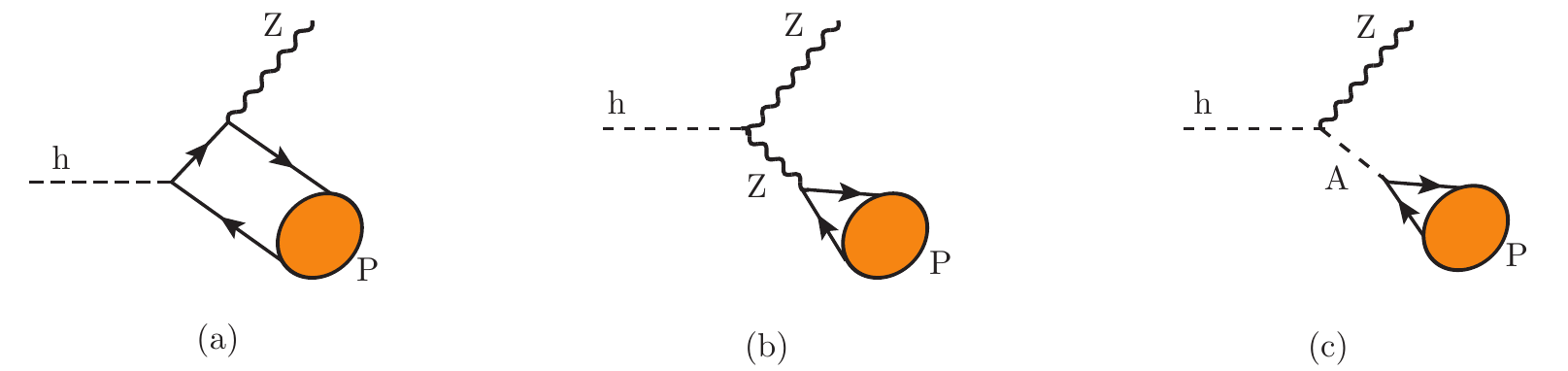}
  \caption{ \sl Diagrams contributing $h\to PZ$ decay.}
  \label{fig:fey1}
\end{figure}
In the SM the decay $h\to PZ$ occurs via the diagrams (a) and (b) shown in Fig.~\ref{fig:fey1}. The dominant (``indirect") contribution comes from $h\to Z Z^\ast \to Z P$ (Fig.~\ref{fig:fey1}b) which is much larger than the direct one, $ h \to Z\bar qq\to Z P$ (Fig.~\ref{fig:fey1}a). In a 2HDM an additional significant contribution arises from $h\to Z A^\ast \to Z P$ (Fig.~\ref{fig:fey1}c), which can be large if the CP-odd Higgs is light and the corresponding couplings to fermions are non-negligible. The decay amplitude then can be written as,
\bea\label{eq:FMPZ}
\mathcal{M}(h\to PZ ) = \frac{g}{v\cos\theta_W} \ \left( k\cdot \varepsilon_Z^\ast \right) \ F^\mathrm{PZ}, 
\eea
with 
\bea\label{eq:FPZ}
F^\mathrm{PZ}\approx {m_Z^2\over m_Z^2 - m_P^2} f_P g_A^q - {f_P v\over m_A^2 - m_P^2 + i m_A\Gamma_A} {m_P^2\over 2 m_q} {m_q\over v}\xi_A^q\cos(\beta-\alpha ), 
\eea
where we displayed only the dominant SM contribution and the one arising from the CP-odd Higgs in the 2HDM. In the above formula $g_A^q=T_q^3$, $\xi_A^q$ is a coupling of the $q\bar q$-par  to the CP-odd Higgs state which will be defined below [cf. Eq.~\eqref{eq:hHA_couplings}], 
and we have used the standard definition of the decay constant $f_P$, namely,
\bea
\langle P(k)\vert \bar q\gamma^\mu \gamma_5 q\vert 0\rangle = - i f_P k_\mu,\qquad 
\langle P(k)\vert \bar q\gamma^5 q\vert 0\rangle = - i f_P {m_P^2\over 2 m_q},
\eea
where $q=c$ or $b$, for $P=\eta_c$ or $\eta_b$, respectively. We should emphasize that owing to the fact that  $\eta_{c,b}$ is a flavor singlet and therefore it can also couple to $(
\alpha_s/4\pi) F_{\mu\nu} \widetilde F^{\mu\nu}$, with $F$ ($\widetilde F$) being the (dual) QCD field strength tensor. While such a term is important for the light quark flavor singlets ($\pi^0$, $\eta$, $\eta^\prime$), it is expected to be very small for the case of heavy quarkonia, which is why it will be neglected in the following.~\footnote{This can be understood by noting that heavy quarkonia are compact hadronic states, $r\sim (m_{c,b} v)^{-1}$, so that their overlap with soft gluons $\sim \Lambda_\mathrm{QCD}^{-1}$ should be very small.}
With the above definitions, the decay rate then reads,
\bea
\Gamma(h\to PZ ) = {\lambda^{3/2}(m_h,m_P,m_Z)\over 16\pi v^4 m_h^3} \vert F^\mathrm{PZ}\vert^2\,,
\eea
where $\lambda(a,b,c)=[a^2-(b+c)^2][a^2-(b-c)^2]$. Notice that in the above amplitude we kept the width of the CP-odd Higgs boson different from zero. In the limit in which $\Gamma_A/m_A\ll 1$, one can work in the narrow width approximation, which amounts to replacing 
\bea
{1\over (q- m_X)^2 + m_X^2\Gamma_X^2} \to \delta(q^2-m_X^2) {\pi \over m_X \Gamma_X} \, .
\eea  
This approximation is adopted throughout this paper both for $X=Z$ and $X=A$.  
We emphasize that, for clarity, we disregarded the direct contributions to the amplitude because they are numerically much smaller and,  since the corresponding expression 
is more complicated, we decided to relegate it to the Appendix.~\footnote{We checked that the direct contribution is indeed numerically much smaller than the first term 
in Eq.~\eqref{eq:FPZ}, and that the decay rate becomes larger by no more than $7\%$ when it is included in the calculation. }
To get the branching fraction in a 2HDM setup, one should be particularly careful with the width of the Higgs boson $\Gamma_h$ which should not be much larger than its SM value, e.g. $\Gamma_h/\Gamma_h^\mathrm{SM}\lesssim 1.4$, a condition that provides a particularly stringent bound on the coupling of $h$ to two light CP-odd Higgses in the situation in which 
$m_A \leq m_h/2$. To be more specific, from the general 2HDM potential, we can read off the term corresponding to the trilinear interaction, namely,
\bea
\mathcal{L}_\mathrm{2HDM} \supset \frac{v}{2!}\lambda_{hAA}\  h AA,
\eea
where $v\approx 246$~GeV, and the coupling
\begin{align}\label{eq:laa}
\lambda_{hAA}= \frac{1}{2 v^2 \sin(2\beta)} & \biggl\{ m_h^2 \left[ \cos(\alpha-3\beta ) + 3 \cos(\alpha +\beta) \right] \biggr.\nn\\
& \biggl. \quad + 4 m_A^2\sin(2\beta)\sin(\beta -\alpha) - 4 M^2 \cos(\alpha+\beta)\biggr\}\,,
\end{align}
with $\alpha,~\beta$ and $M^2$ explicitly defined in Sec.~\ref{sec:2hdm}. The expression for the decay width reads:
\bea
\Gamma(h\to AA)=  {|\lambda_{hAA}|^2\over 32\pi } \frac{v^2}{ m_h} \sqrt{1- \frac{4m_A^2}{m_h^2}}\,.
\eea
Besides $h\to AA$, if opened, the channel $h\to ZA$ can give significant contribution to the width $\Gamma_h$. The decay width of this particular channel is given by 
\bea
\Gamma(h\to ZA)=  {1\over 16\pi } \frac{\cos^2(\beta-\alpha)}{m_h^3 v^2} \lambda^{3/2}(m_h,m_Z,m_A) \,.
\eea
\subsection{$\cb (h\to P\ell^+\ell^-)$}

\begin{figure}[h!]
  \centering
  \includegraphics[width=\textwidth]{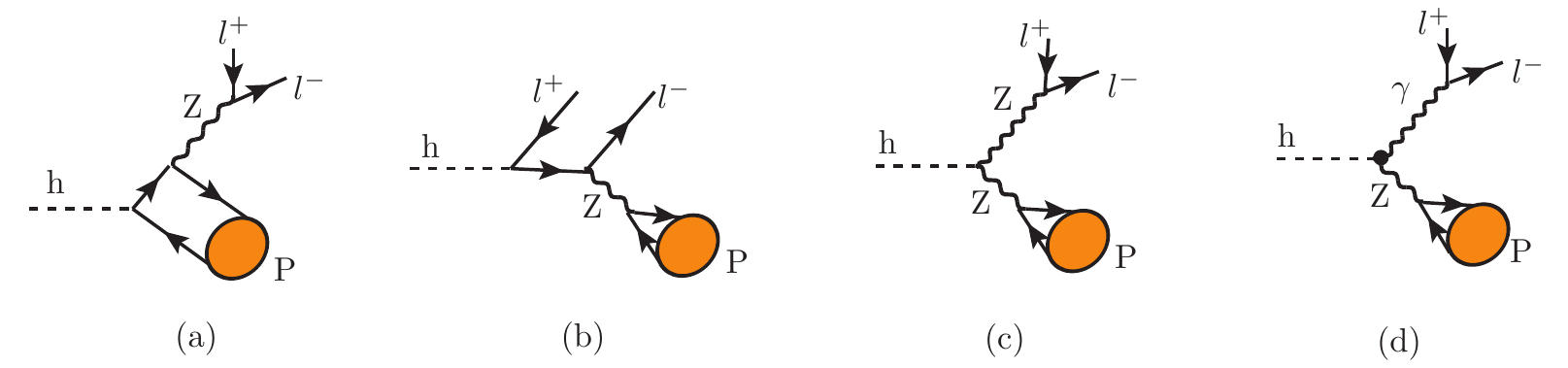}
  \caption{ \sl Diagrams relevant to $h\to P\ell^+\ell^-$ decay in the Standard Model. The full dot in the diagram (d) indicates that the vertex is loop-induced. 
  Its contribution to the decay rate is nevertheless zero.}
  \label{fig:fey2}
\end{figure}

In the Standard Model the situation with $h\to P\ell^+\ell^-$ is similar to the one discussed in the case of $h\to PZ$. 
The dominant contribution comes from the diagram Fig.~\ref{fig:fey2}c
which reads
\begin{align}\label{eq:fig2c}
\mathcal{M}(h\to P\ell^+\ell^- )^\mathrm{2c} =&-\frac{1}{4} \left(\frac{g}{\cos\theta_W}\right)^3 m_Z\ {g_A^q f_P \over (q^2 - m_Z^2)\ (k^2-m_Z^2)}\left(-g^\alpha_\mu+{q^\mu q_\alpha\over m_Z^2}\right) \nn\\
&\left(-g^{\nu\alpha}+{k^\nu k^\alpha\over m_Z^2}\right) \ k_\nu\ \bar u_\ell \gamma^\mu (g_V^\ell - g_A^\ell \gamma_5) v_\ell\,.
\end{align}
The full expression, which includes all contributions depicted in Fig.~\ref{fig:fey2}, is given in Appendix of the present paper. We note, however, that the contribution of the diagram in 
Fig.~\ref{fig:fey2}d vanishes in the decay rate.

\begin{figure}[h!]
  \centering
  \includegraphics[width=\textwidth]{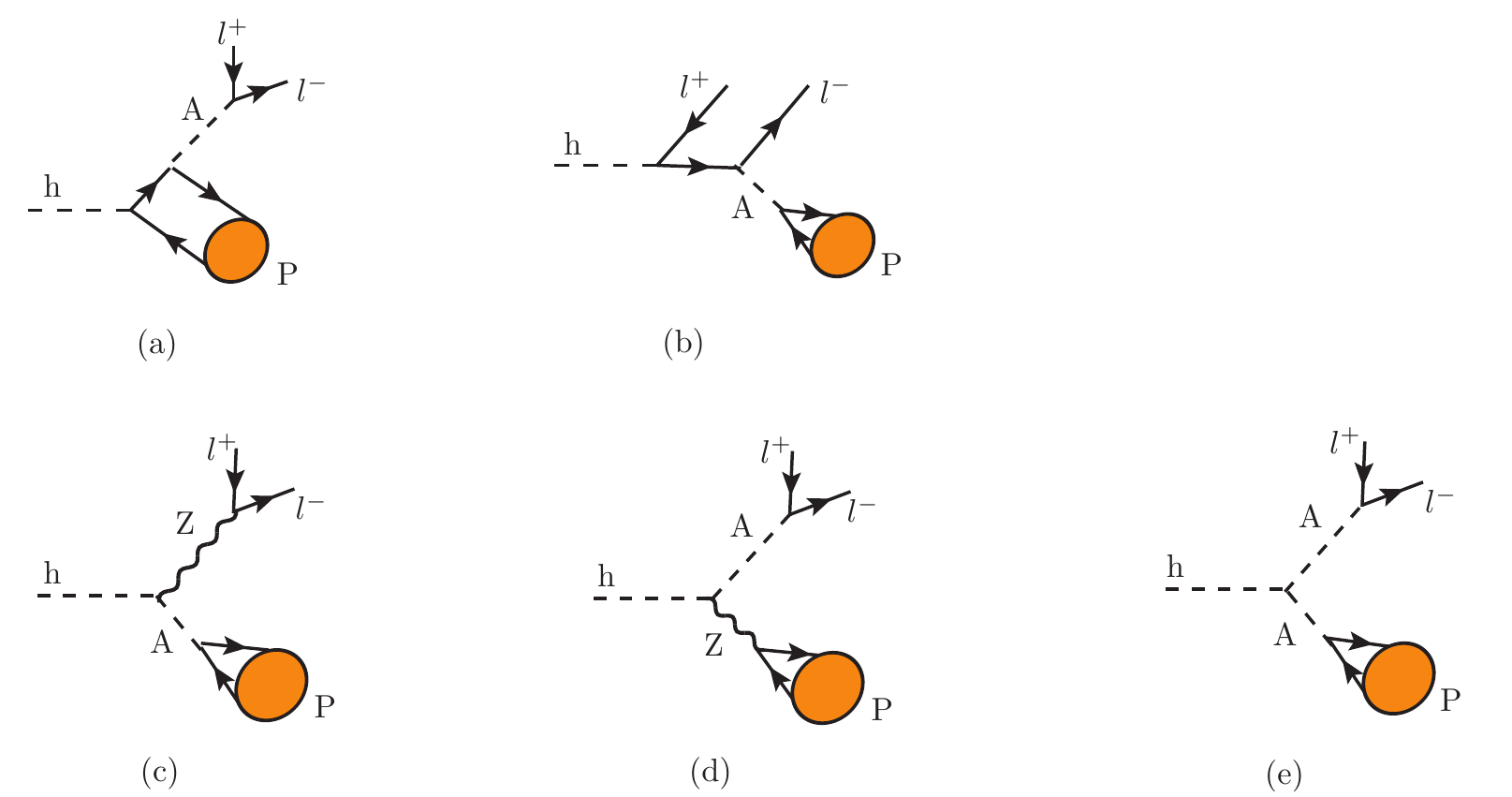}
  \caption{ \sl Contributions to the $h\to P\ell^+\ell^-$ decay amplitude in a 2HDM scenario. }
  \label{fig:fey3}
\end{figure}

Diagrams arising in the 2HDM setup are shown in Fig.~\ref{fig:fey3}, of which the first two are numerically much less significant than the remaining three. The contributions of those latter (dominant) diagrams to the decay amplitude read:
 \begin{align}
\mathcal{M}(h\to P\ell^+\ell^- )^\mathrm{3c} =&-\left(\frac{g}{2\cos\theta_W}\right)^2\frac{m_q \xi^q_A}{v} \frac{m_P^2 f_P }{2 m_q} \ {\cos(\beta-\alpha ) \over (q^2 - m_Z^2)\ (k^2-m_A^2)}  \nn\\
&\left(-g_{\mu \nu }+{q_\mu q_\nu \over m_Z^2}\right) \ (k+p)^\mu\ \bar u_\ell \gamma^\nu (g_V^\ell - g_A^\ell \gamma_5) v_\ell\,, \label{eq:fig3c} \\
\mathcal{M}(h\to P\ell^+\ell^- )^\mathrm{3d} =&\left(\frac{g}{2\cos\theta_W}\right)^2\frac{m_\ell \xi^\ell_A}{v}  {g_A^q f_P \cos(\beta-\alpha ) \over (q^2 - m_A^2)\ (k^2-m_Z^2)}  \nn\\
&\left(-g_{\mu \nu }+{k_\mu k_\nu \over m_Z^2}\right) \ (q+p)^\mu k^\nu\ \bar u_\ell  \gamma_5 v_\ell\,,  \label{eq:fig3d} \\
\mathcal{M}(h\to P\ell^+\ell^- )^\mathrm{3e} =-&\lambda_{hAA} v\ \frac{m_q \xi^q_A}{v} \frac{m_\ell \xi^\ell_A}{v}  \frac{m_P^2 f_P }{2 m_q}   {1\over (q^2 - m_A^2)\ (k^2-m_A^2)}    \bar u_\ell  \gamma_5 v_\ell\, \label{eq:fig3e}. 
\end{align}
To a very good approximation the decay rate can be written as
\bea\label{eq:total}
\Gamma( h\to P\ell^+\ell^- ) \simeq \Gamma( h\to PZ^\ast \to P \ell^+\ell^- ) + \Gamma( h\to PA^\ast \to P \ell^+\ell^- ) .
\eea
We checked that the interference terms are indeed very small and the above formula is useful for the phenomenology we are interested in. For the explicit expressions of the separate rates in Eq.~\eqref{eq:total} 
we obtain,
\begin{align}
\Gamma( h\to PZ^\ast \to P \ell^+\ell^- )=& { f_P^2 m_Z^3\over 384 \pi^2 \Gamma_Z m_h^3 v^6 } \bigl[ \cos^2(2\theta_W) + 4\sin^4\theta_W\bigr] \nn\\
& \left( g_A^q- {\xi_A^q m_P^2\cos(\beta-\alpha)\over 2(m_A^2-m_P^2) } \right)^2 \lambda^{3/2}(m_h,m_P,m_Z),\\
\Gamma( h\to PA^\ast \to P \ell^+\ell^- )=& { f_P^2 m_A\over 512 \pi^2 \Gamma_A m_h^3 v^2} \left( \frac{m_\ell \xi_A^\ell}{v}  \right)^2  \biggl[ \lambda_{hAA} \frac{m_P^2}{m_A^2-m_P^2} \frac{\xi_A^q}{v} v^2  \biggr. \nn\\
&\biggl.+ 2\cos(\beta-\alpha){g_A^q\over v} (m_h^2-m_A^2)   \biggr]^2\ \lambda^{1/2}(m_h,m_P,m_A).
\end{align}
In the above formulas $g_V^f=T_f^3- 2Q_f\sin^2\theta_W$, $g_A^f=T_f^3$, and we neglected the additive terms $\propto m_\ell^2/m_Z^2$. 
We emphasize once again that the above expression for $\Gamma( h\to P\ell^+\ell^- )$ should be viewed as a very good approximation, whereas the full expressions are provided in the Appendix 
of the present paper.  

Finally, since the above formulas require the knowledge of the width of $A$, we give that expression too, in which we include $\Gamma_A = \Gamma(A\to f\bar f) +   \Gamma(A\to \gamma \gamma )$, 
which are given by
\begin{align}
\Gamma(A\to f\bar f)  &= \theta(m_A-2m_f) |\xi_A^f|^2\, {N_c m_f^2\over 8\pi v^2}\, m_A\, \sqrt{1-\frac{4m_f^2}{m_A^2}} \,,\nn \\ 
\Gamma(A\to \gamma \gamma ) &= \frac{ \alpha_\mathrm{em}^2 }{16 \pi^3 v^2} m_A^3 \left| \sum_f \xi_A^f \, N_c \, Q_f^2 \, x_f\, F_{\gamma\gamma}(x_f) \right|^2 \,,
\end{align}
where $N_c=3$ for quarks and $N_c=1$ for leptons, the electric charge for leptons, up-type and down-type quarks is $Q_\ell =-1$, $Q_u = 2/3$,  $Q_d =-1/3$, respectively, and $f$ runs over all available quark and lepton flavors. In the above expression we used the notation $x_f=m_f^2/m_A^2$, and the loop function reads 
\begin{equation}
F_{\gamma\gamma}(x) = 
\left\{\begin{array}{lr}
 \frac{1}{2} \left( i \pi + \log\left[ \displaystyle{1+\sqrt{1-  4 x} \over 1-\sqrt{1-  4 x}}\right] \right)^2, & \text{for } x< 1/4\\
 - 2 \arcsin^2\left( \displaystyle{1\over 2 \sqrt{x} }\right),  & \text{otherwise }  
        \end{array}\right. .
\end{equation} 
Notice that $\Gamma(A\to \gamma \gamma ) \ll \Gamma(A\to f\bar f)$ and our conclusion would remain the same even if we neglected $\Gamma(A\to \gamma \gamma )$. For the same reason we neglect the $\Gamma(A\to gg )$ contribution to the full width of $A$.

\section{Scanning the 2HDM parameter space}
\label{sec:2hdm}

We consider a general CP-conserving 2HDM with a softly broken $\mathbb{Z}_2$ symmetry, described by the following potential:
\begin{align}
\label{eq:V2hdm}
V(\Phi_1,\Phi_2) = m_{11}^2\Phi_1^\dagger\Phi_1 & +m_{22}^2\Phi_2^\dagger\Phi_2 + m_{12}^2(\Phi_1^\dagger\Phi_2+\Phi_2^\dagger\Phi_1)+\dfrac{\lambda_1}{2}(\Phi_1^\dagger \Phi_1)^2+\dfrac{\lambda_2}{2}(\Phi_2^\dagger \Phi_2)^2\nonumber\\
&+\lambda_3 \Phi_1^\dagger\Phi_1 \Phi_2^\dagger\Phi_2+\lambda_4 \Phi_1^\dagger\Phi_2 \Phi_2^\dagger\Phi_1+\dfrac{\lambda_5}{2}\left[ (\Phi_1^\dagger\Phi_2)^2+(\Phi_2^\dagger\Phi_1)^2\right],
\end{align}
where the term proportional to $m_{12}^2$ breaks the $\mathbb{Z}_2$ symmetry. In the above expression $\Phi_a$ ($a=1,2$) stand for a scalar doublet
\begin{align}\label{eq:higgs0}
\Phi_a(x) = \begin{pmatrix}
\phi_a^+(x) \\ 
\frac{1}{\sqrt{2}}(v_a+\rho_a(x)+i \eta_a(x))
\end{pmatrix}, 
\end{align}
where $v_{1,2}$ are the vacuum expectation values related to $v^\mathrm{SM}=\sqrt{v_1^2+v_2^2}=246.22$~GeV~\cite{Olive:2016xmw}. 
Two of the above fields are Goldstone bosons ($G^0$, $G^\pm$), two are massive CP-even states ($h$, $H$), one CP-odd ($A$), 
and one charged Higgs ($H^\pm$). These fields are related to the ones given in Eq.~\eqref{eq:higgs0} via 
\begin{align}
\begin{pmatrix}
\phi_1^+\\ \phi_2^+
\end{pmatrix}
&=\begin{pmatrix}
\cos \beta & - \sin \beta\\
\sin\beta  & \cos \beta
\end{pmatrix}
\begin{pmatrix}
G^+\\ H^+
\end{pmatrix},\qquad
\begin{pmatrix}
\eta_1\\ \eta_2
\end{pmatrix}
=\begin{pmatrix}
\cos \beta & - \sin \beta\\
\sin\beta  & \cos \beta
\end{pmatrix}
\begin{pmatrix}
G^0\\ A
\end{pmatrix},\nn
\end{align}
\begin{align}
\begin{pmatrix}
\rho_1\\ \rho_2
\end{pmatrix}
&=\begin{pmatrix}
\cos \alpha & - \sin \alpha\\
\sin\alpha  & \cos \alpha
\end{pmatrix}
\begin{pmatrix}
H\\ h
\end{pmatrix},
\end{align}
where the mixing angles $\alpha$ and $\beta$ satisfy
\begin{align}
\tan \beta = \frac{v_2}{v_1},\qquad \tan 2\alpha = \dfrac{2(-m_{12}^2+\lambda_{345}v_1 v_2)}{m_{12}^2(v_2/v_1-v_1/v_2)+\lambda_1 v_1^2-\lambda_2 v_2^2},
\end{align}
with $\lambda_{345}\equiv \lambda_3+\lambda_4+\lambda_5$.
The masses of the physical scalars can be written in terms of parameters which appear in the potential as
\begin{align}
\label{eq:massesH}
	m_H^2 &= M^2 \sin^2(\alpha-\beta)+\left(\lambda_1 \cos^2\alpha \cos^2\beta+\lambda_2 \sin^2\alpha \sin^2\beta+\frac{\lambda_{345}}{2}\sin 2\alpha \sin 2\beta\right)v^2,\nn\\
	m_h^2 &= M^2 \cos^2(\alpha-\beta)+\left(\lambda_1 \sin^2\alpha \cos^2\beta+\lambda_2 \cos^2\alpha \sin^2\beta-\frac{\lambda_{345}}{2}\sin 2\alpha \sin 2\beta\right)v^2,\nn\\
	m_{A}^2 &= M^2-\lambda_5 v^2,\nn\\
	m_{H^\pm}^2 &= M^2-\frac{\lambda_{4}+\lambda_5}{2} v^2,
\end{align}
where for shortness we use $M^2= m_{12}^2/(\sin \beta \cos \beta)$. As far as the Yukawa sector is concerned, the $\mathbb{Z}_2$ symmetry is imposed to prevent the flavor changing 
processes to appear at tree level~\cite{Glashow:1976nt} by enforcing each type of the right-handed fermion to couple to a single Higgs doublet. Four choices are then possible and they 
are called Type I, II, X and Z~\cite{Branco:2011iw}. By writing the Yukawa Lagrangian as  
\begin{align}
\label{eq:lyuk}
\mathcal{L}_Y = &- \dfrac{\sqrt{2}}{v} H^+ \Big{\lbrace} \bar{u}~[\zeta^d \, V_\mathrm{CKM} m_d P_R-\zeta^u \, m_u V_\mathrm{CKM} P_L]~d +\zeta^\ell \, \bar{\nu} m_\ell P_R \ell \Big{\rbrace}\nonumber \\
&-\dfrac{1}{v}\sum_{f,\varphi_i^0\in\{h,H,A\}} \xi^f_{\varphi_i^0}~ \varphi_i^0 \, \Big{[}\bar{f} m_f P_R f \Big{]}+\mathrm{h.c.},
\end{align}
where a specific choice of parameters $\zeta^f$ corresponds to the above mentioned  types of 2HDM, cf. Table \ref{tab:y2hdm}. In the expression $u$, $d$ and $\ell$ stand for the up-type, down-type quark, and a lepton flavor respectively, and $f$ for a generic fermion. $V_\mathrm{CKM}$ is the Cabibbo--Kobayashi-Maskawa matrix, lepton mixing are neglected, and $P_{L,R}= (1\mp\gamma_5)/2$. 
Furthermore, the couplings $\xi^f_{\varphi_i^0}$ are related to $\zeta^f$ as:
\begin{align}\label{eq:hHA_couplings}
	\xi_h^f &= \sin(\beta-\alpha) +\cos(\beta-\alpha) \zeta^f, \nonumber \\
	\xi_H^f &= \cos(\beta-\alpha) -\sin(\beta-\alpha) \zeta^f, \nonumber \\
	\xi_A^{u}&=-i\zeta^u,\qquad \xi_A^{d,\ell}=i \zeta^{d,\ell}.
\end{align}
\begin{table}[ht!]
\renewcommand{\arraystretch}{1.5}
\centering
\begin{tabular}{|c|c|c|c|}
\hline 
Model & $\zeta^d$ & $\zeta^u$ & $\zeta^\ell$ \\ \hline
Type I &\quad $\cot \beta$ \quad & \quad $\cot \beta$ \quad & \quad $\cot \beta$ \quad\\  
Type II & $-\tan \beta$  &  $\cot \beta$ & $-\tan \beta$ \\  
Type X (lepton specific) & $\cot \beta$  &  $\cot \beta$ & $-\tan \beta$  \\
Type Z (flipped) & $-\tan \beta$  &  $\cot \beta$ & $\cot \beta$ \\
  \hline
\end{tabular}
\caption{ \sl Couplings $\zeta^f$ in various types of 2HDM.}
\label{tab:y2hdm} 
\end{table}

Once we spelled out all the parameters of 2HDM we need to perform a scan of the parameters by taking into account the following general theory constraints:
\begin{itemize}
\item \textbf{Stability} of the scalar potential is ensured by the requirement that it is bounded from below, which is achieved if the quartic couplings satisfy~\cite{Gunion:2002zf}
	\begin{equation}
	\lambda_{1,2}>0,\qquad \lambda_3>-(\lambda_1 \lambda_2)^{1/2},\qquad  \lambda_3+\lambda_4- \vert \lambda_5 \vert  >-(\lambda_1 \lambda_2)^{1/2},
\end{equation}		
while the stability of the electroweak vacuum requires
\begin{align}
m_{11}^2+\dfrac{\lambda_1 v_1^2}{2}+\dfrac{\lambda_3 v_2^2}{2} &= \frac{v_2}{v_1} \left[ m_{12}^2 - (\lambda_4+\lambda_5)\dfrac{v_1 v_2}{2}\right],\\
m_{22}^2+\dfrac{\lambda_2 v_2^2}{2}+\dfrac{\lambda_3 v_1^2}{2} &= \frac{v_1}{v_2} \left[ m_{12}^2 - (\lambda_4+\lambda_5)\dfrac{v_1 v_2}{2}\right].
\end{align} 
In order for a minimum to be global, the parameters should also satisfy~\cite{Barroso:2013awa}:
\begin{equation}
m_{12}^2 \left(m_{11}^2-m_{22}^2 \sqrt{\lambda_1/\lambda_2} \right) \left( \tan \beta - \sqrt[4]{\lambda_1/\lambda_2}\right) >0.
\end{equation} 

\item \textbf{Unitarity} of the $S$-wave of the scalar scattering amplitudes gives rise to the following inequalities~\cite{Kanemura:1993hm,Swiezewska:2012ej}
\begin{equation}
|a_\pm|, |b_\pm|, |c_\pm|, |f_\pm|, |e_{1,2}|, |f_1|, |p_1| < 8 \pi,
\end{equation}
where
\begin{align}
\begin{split}
a_\pm &= \dfrac{3}{2}(\lambda_1+\lambda_2)\pm \sqrt{\dfrac{9}{4}(\lambda_1-\lambda_2)^2+(2\lambda_3+\lambda_4)^2},\\
b_\pm &= \dfrac{1}{2}(\lambda_1+\lambda_2)\pm \dfrac{1}{2} \sqrt{(\lambda_1-\lambda_2)^2+4\lambda_4^2},\\
c_\pm &= \dfrac{1}{2}(\lambda_1+\lambda_2)\pm \dfrac{1}{2} \sqrt{(\lambda_1-\lambda_2)^2+4\lambda_5^2},\\
e_1 &= \lambda_3 + 2 \lambda_4 -3\lambda_5,\hspace*{3cm}
e_2 = \lambda_3-\lambda_5,\\
f_+ &= \lambda_3+2 \lambda_4+3\lambda_5, \hspace*{2.9cm} f_- =\lambda_3+\lambda_5,\\
f_1 &= \lambda_3+\lambda_4, \hspace*{4.3cm}p_1 = \lambda_3-\lambda_4.
\end{split}
\end{align}
\item \textbf{Electroweak precision tests} provide important constraint to the 2HDM parameters. We consider the expressions for the parameters 
$S$, $T$ and $U$ in 2HDM~\cite{Barbieri:2006bg}, and the values for  $\Delta S$, $\Delta T$, $\Delta U$ given in Ref.~\cite{Baak:2014ora},
\begin{equation}
  \begin{aligned}
    & \Delta S^{\rm SM} = 0.05\pm 0.11, \\
    & \Delta T^{\rm SM} = 0.09\pm 0.13, \\
    & \Delta U^{\rm SM} = 0.01\pm 0.11, \\
  \end{aligned}
\qquad\qquad
\mathrm{cov} = \left(\begin{array}{ccc}
1 & 0.90 & -0.59 \\
0.90 & 1 & -0.83 \\
-0.59 & -0.83 & 1
\end{array}\right), 
\end{equation}
to $99\%$ C.L. Using $X=(\Delta S, \Delta T, \Delta U)$, $\sigma =(0.11,0.13,0.11)$, and $\sigma_{ij}^2\equiv \sigma_i \mathrm{cov}_{ij} \sigma_j$, we build
\begin{equation}
  \chi^2= \sum_{i,j}(X_i - X_i^{\rm SM})(\sigma^2)_{ij}^{-1}(X_j - X_j^{\rm SM}).
\end{equation}
and choose the points which agree with the above numbers.

\end{itemize}

\begin{figure}[ht]
  \centering
  \includegraphics[width=0.5\textwidth]{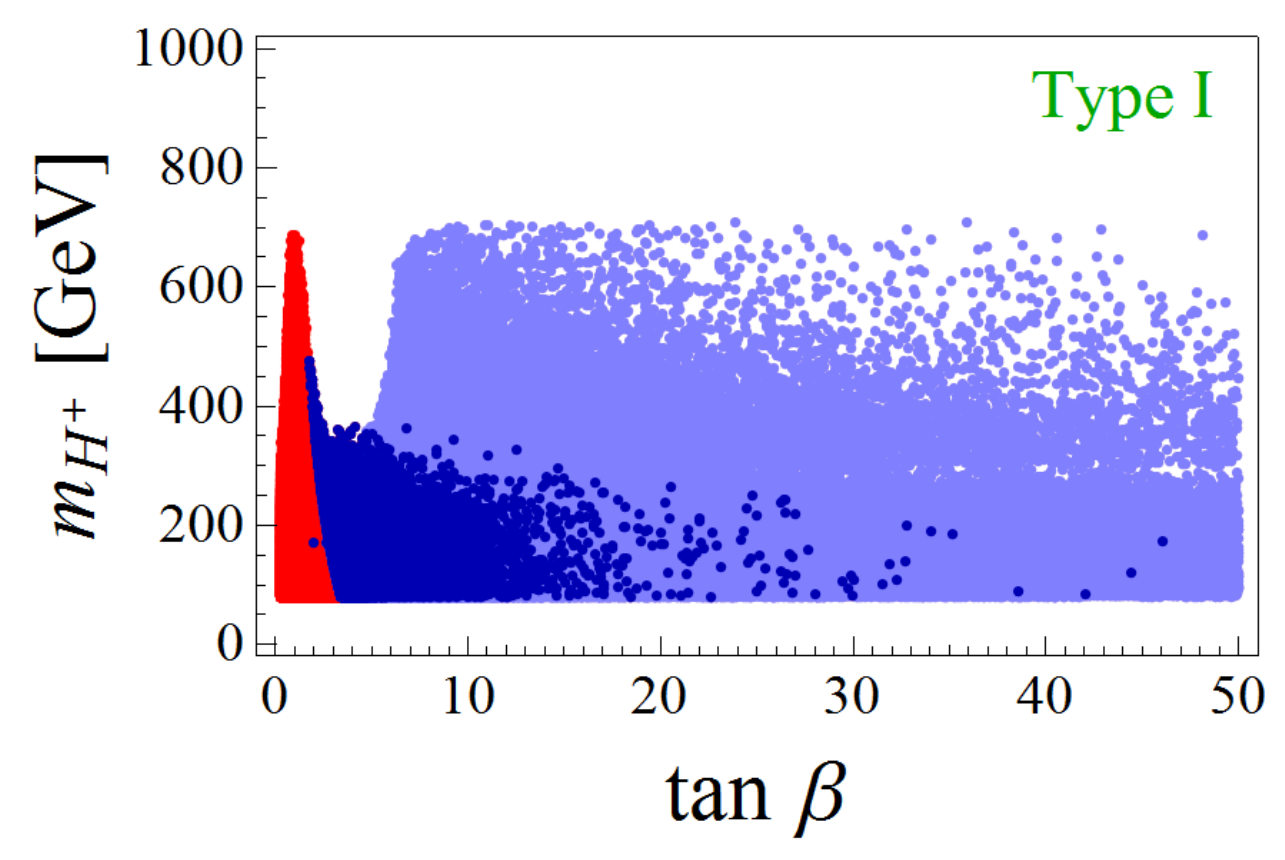}~\includegraphics[width=0.5\textwidth]{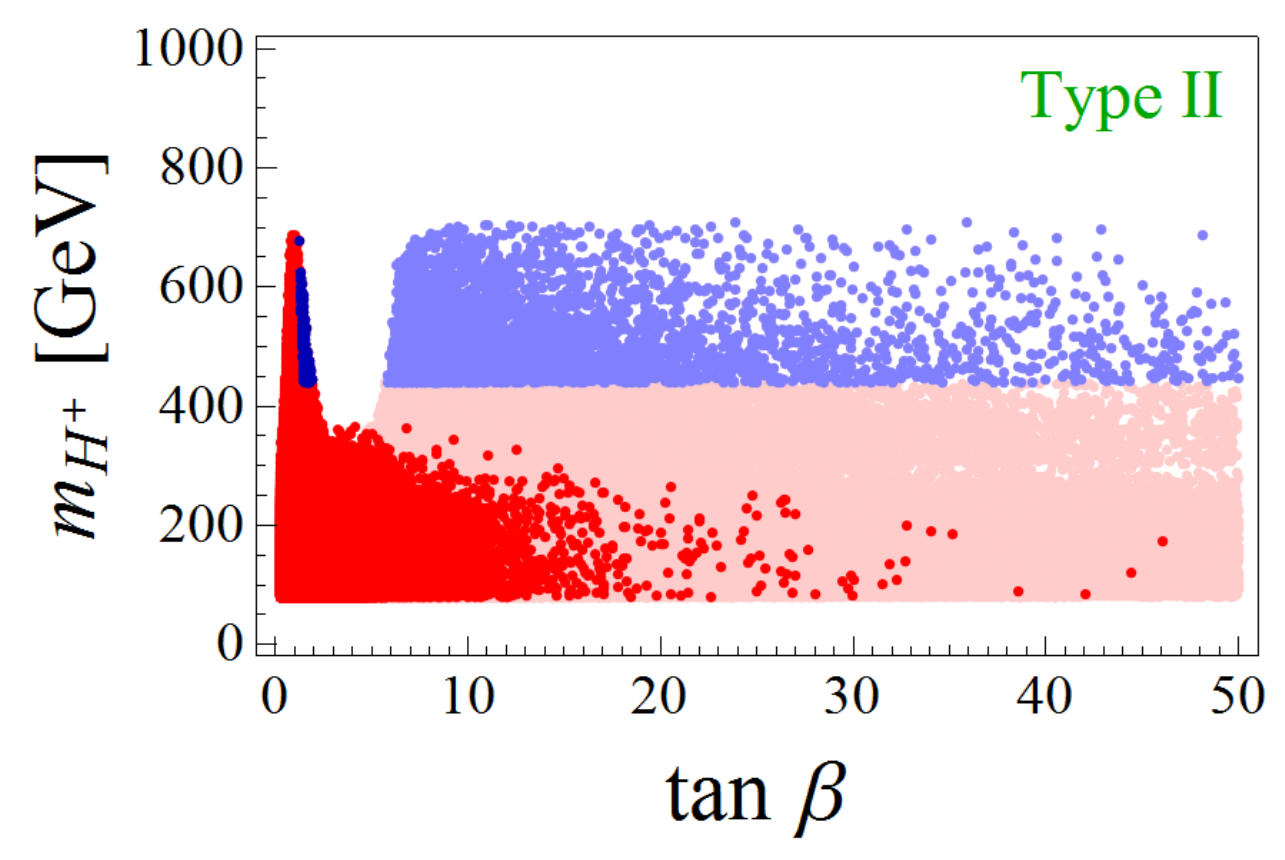}  \\
  \includegraphics[width=0.5\textwidth]{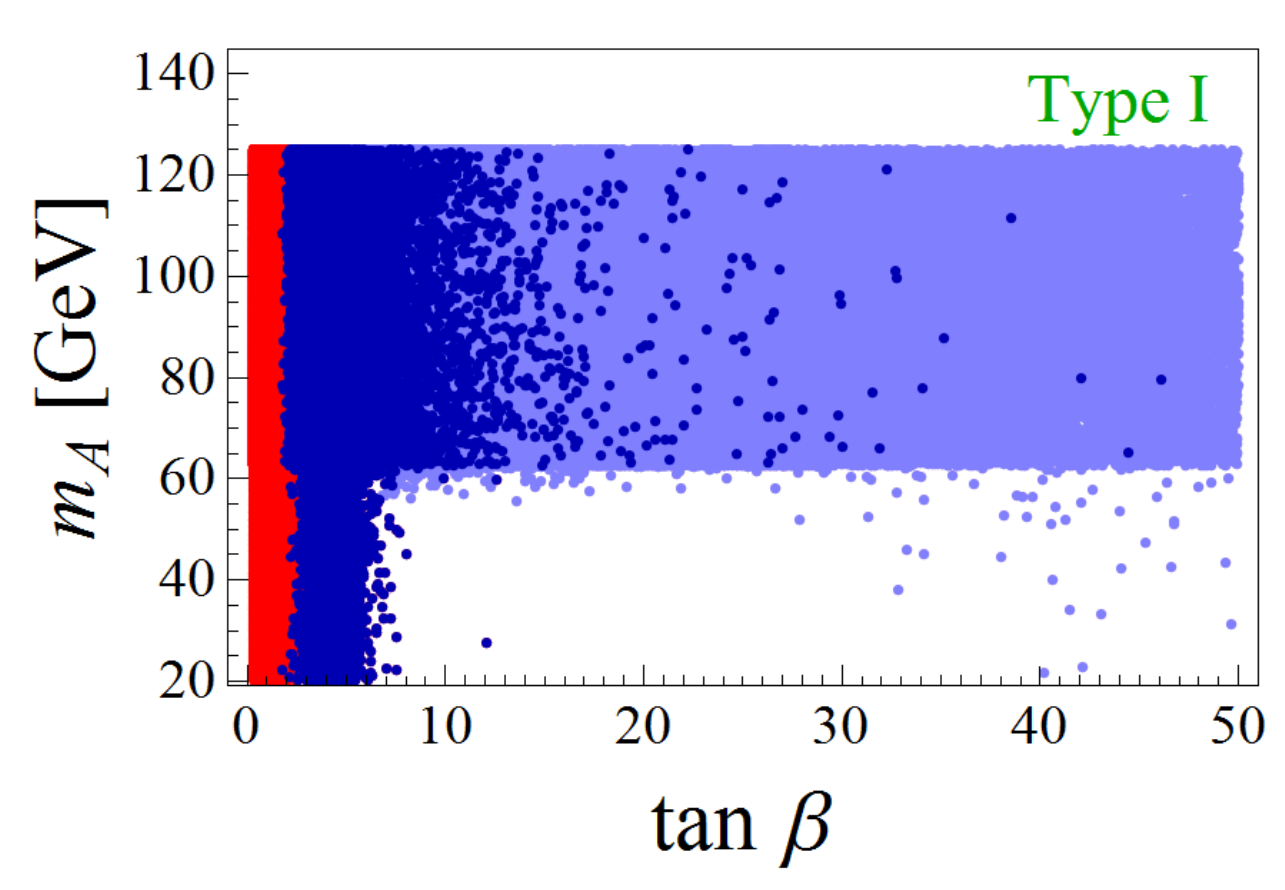}~\includegraphics[width=0.5\textwidth]{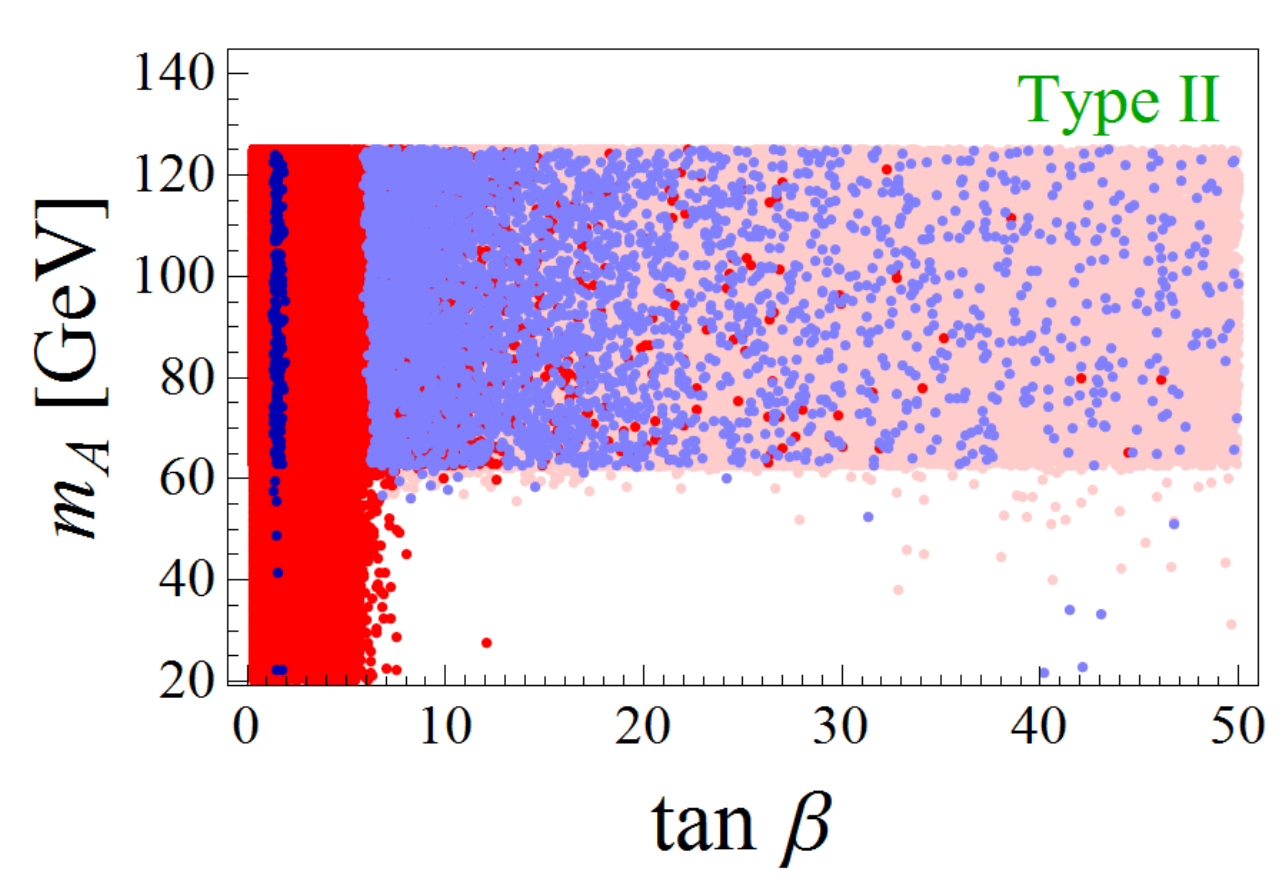}
  \caption{ \sl Results of the scan of parameters~\eqref{eq:scan2} after imposing constraints discussed in the text. Darker/lighter points correspond to the {\sl free}/{\sl fine-tuned} scan. 
Notice in particular that the red points are forbidden by the flavor bounds~\cite{Misiak:2017bgg,Arnan:2017lxi}.} 
  \label{fig:scan}
\end{figure}

In our scan we identify the lightest CP-even state ($h$) with the SM-like scalar, observed at the LHC, with mass $m_h=125.09(24)$~GeV~\cite{Olive:2016xmw}. We assume $\cb(h\to \mathrm{invisible})\leq 0.3$, and impose the near-alignment condition, $|\cos(\beta-\alpha)|\leq 0.3$, in order to ensure that $g_{hWW}$ and $g_{hZZ}$ remain consistent 
with the measured values and in agreement with the SM predictions~\cite{Corbett:2015ksa}~\footnote{
By $\cb(h\to \mathrm{invisible})\leq 0.3$ we mean the contribution to the width $\Gamma_h$ arising from the 2HDM alone. In other words $\cb(h\to \mathrm{invisible}) = (\Gamma_h - \Gamma_h^\mathrm{SM})/\Gamma_h$, so that the 
requirement $\cb(h\to \mathrm{invisible})\leq 0.3$ could be interpreted as $\Gamma_h/\Gamma_h^\mathrm{SM}\leq 1.42$.
}. The above-mentioned constraints are then imposed onto a set of randomly generated points in the intervals: 

\begin{align}\label{eq:scan2}
\begin{split}
	&\tan \beta \in (0.2,50),\qquad \hspace*{1.5cm}\alpha\in\left(-\frac{\pi}{2},\frac{\pi}{2}\right), \qquad \hspace*{1cm} \left|M^2\right| \leq (1~\mathrm{TeV})^2,\\[0.6em]
	&m_{H^\pm}\in (m_W, 1~\mathrm{TeV}),\qquad m_{H}\in (m_h, 1~\mathrm{TeV}),\qquad m_{A}\in \left(20\ \gev, m_h\right).
\end{split}
\end{align}
Due to correlation between $m_A$ and $m_{H^\pm}$, the fact that we choose to work with light CP-odd Higgs implies that the charged Higgs is bounded from above. We find $m_{H^\pm} \lesssim 700$~GeV, as shown in Fig.~\ref{fig:scan}. In addition to the above constraints we also impose the bound arising from the comparison between theory predictions and the experimental results concerning the spectrum of $B\to X_s\gamma$ decay, which for the case of 
Type~II and Type~Z models amounts to a $3\sigma$ bound of $m_{H^\pm}\geq 439$~GeV~\cite{Misiak:2017bgg}. 
Furthermore, the $3\sigma$ bounds coming from the exclusive $b\to s\mu^+\mu^-$ decay modes, discussed in Ref.~\cite{Arnan:2017lxi}, have also been included.  
%%%%%%%%%%%%%%%%%%%%%%%%%%%%%%%%%%%%%%%%%%%%%%
%%%%%%%%%%%%%%%%%%%%%%%%%%%%%%%%%%%%%%%%%%%%%%
%%%%%%%%%%%%%%%%%%%%%%%%%%%%%%%%%%%%%%%%%%%%%%
%%%%%%%%%%%%%%%%%%%%%%%%%%%%%%%%%%%%%%%%%%%%%%
%%%%%%%%%%%%%%%%%%%%%%%%%%%%%%%%%%%%%%%%%%%%%%
%%%%%%%%%%%%%%%%%%%%%%%%%%%%%%%%%%%%%%%%%%%%%%
It is important to mention that a scan of parameters consistent with the constraints listed above favors the moderate and small values of $\tan\beta \in (0.2,15]$. 
Larger values of $\tan \beta$ can be probed around the alignment limit as it can be seen from Eq.~\eqref{eq:massesH}. For that reason, in addition to the {\sl free} scan, we also perform a {\sl fine-tuned} scan, i.e. with $m_H\approx|M|$. 
We combine both results and show them in Fig.~\ref{fig:scan} for Type~I and Type~II models. Similar results are obtained for Type~X and Type~Z. The lower bound on the charged Higgs, $m_{H^\pm}\geq 439$~GeV~\cite{Misiak:2017bgg}, 
eliminates many points in Type~II and Type~Z models.  As it can be seen from Fig.~\ref{fig:scan}, among the remaining points in the parameter space of the Type~II model, the free scan prefers low $\tan\beta \approx 2$ values, while the large 
$\tan\beta$ values can only be accessed through the scan in the $m_H\approx|M|$ direction.  The same observation applies to the Type~Z model. 

Notice also that one cannot simultaneously access $m_A < m_h/2$ and have large $\tan\beta$, except for the very fine tuned solutions. To see that we expand Eq.~\eqref{eq:laa} around the alignment limit, $\delta= \cos(\alpha -\beta ) \approx 0$, and obtain
\bea
\lambda_{hAA}= \frac{m_h^2+2m_A^2-2M^2}{v^2} + \frac{2(m_h^2-m_H^2)}{v^2 \tan 2\beta} \delta +\mathcal{O}(\delta^2). 
\eea
To suppress $h\to \mathrm{invisible}$ in the perfect alignment limit, one needs $2M^2  = m_h^2+2 m_A^2$, which for the low $m_A$ gives $M^2\leq 3 m_h^2/4 \ll m_H^2$. Therefore, in order to get the large $\tan\beta$ values one needs  
$M \approx m_H$ which is in contradiction with the previous inequality.

\section{Sensitivity of $\cb (h\to PZ)$ and $\cb (h\to P\ell^+\ell^-)$ on the light CP-odd Higgs state}
\label{sec:eff}
In this Section we use the results of the scan discussed above and evaluate 
\begin{align}\label{eq:RR}
&R^Z_{\eta_{cb}}={\cb (h \to \eta_{cb} Z)^\mathrm{2HDM} \over \cb (h \to \eta_{cb} Z)^\mathrm{SM}}\ , \nn\\
{\phantom{\Huge{l}}}\raisebox{.35cm}{\phantom{\Huge{j}}}
&R^{\tau\tau}_{\eta_{cb}}= {\cb (h \to \eta_{cb} \tau^+\tau^-)^\mathrm{2HDM} \over \cb (h \to \eta_{cb}  \tau^+\tau^-)^\mathrm{SM}} \ , \nn\\
{\phantom{\Huge{l}}}\raisebox{.35cm}{\phantom{\Huge{j}}}
&R^{\mu\mu}_{\eta_{cb}}= {\cb (h \to \eta_{cb} \mu^+\mu^-)^\mathrm{2HDM} \over \cb (h \to \eta_{cb}  \mu^+\mu^-)^\mathrm{SM}}\ ,
\end{align}
where $\eta_{cb}$ means either ${\eta_c}$ or $\eta_b$. Essential hadronic quantities needed to evaluate $\cb (h \to \eta_{cb} Z)^\mathrm{SM}$ and $\cb (h \to \eta_{cb}  \ell^+\ell^-)^\mathrm{SM}$ are~\cite{Becirevic:2013bsa,McNeile:2012qf}:
\bea
f_{\eta_c}= 391\pm 4\ \mev,\qquad f_{\eta_b}= 667\pm 7\ \mev ,
\eea 
which together with masses and constants available in Ref.~\cite{Olive:2016xmw}, combined with the formulas given in Sec.~\ref{sec:h2p}, result in
\begin{align}
\cb(h \to \eta_{c} Z)^\mathrm{SM} &= (1.00\pm 0.01)\times 10^{-5},\nn\\
{\phantom{\Huge{l}}}\raisebox{.15cm}{\phantom{\Huge{j}}}
 \cb(h \to \eta_{b} Z)^\mathrm{SM} &= (2.69\pm 0.05)\times 10^{-5},\nn\\
{\phantom{\Huge{l}}}\raisebox{.15cm}{\phantom{\Huge{j}}}
\cb(h \to \eta_{c} \ell^+\ell^-)^\mathrm{SM} &= (3.40\pm 0.07) \times 10^{-7},\nn\\
{\phantom{\Huge{l}}}\raisebox{.15cm}{\phantom{\Huge{j}}}
 \cb(h \to \eta_{b} \ell^+\ell^-)^\mathrm{SM} &= (8.76\pm 0.06) \times 10^{-7}\,,
\end{align}
where we stress again that the additive (numerically insignificant) terms proportional to $m_\ell^2/m_Z^2$ have been neglected.
After inspection, and by using the result of the scan discussed in the previous Section, the range of values for each of the ratios~\eqref{eq:RR}, that we obtain by using the results of the scan from the previous Section, are summarized in Tab.~\ref{tab:2}.
\begin{table}[ht!]
\renewcommand{\arraystretch}{1.75}
\centering
\begin{tabular}{|c|cc|cc|cc|}
\hline 
Ratio & $R_{\eta_c}^Z$ & $R_{\eta_b}^Z$ & $R_{\eta_c}^{\mu\mu}$ & $R_{\eta_b}^{\mu\mu}$ & $R_{\eta_c}^{\tau\tau}$& $R_{\eta_b}^{\tau\tau}$ \\ \hline
Type I & $(0.7, 1.0)$   & $(0.7,1.0)$ &  $(0.7, 1.0)$        &$(0.7, 1.0)$    &\cellcolor{gray!25} $(0.7, 3.3)$    & \cellcolor{gray!25} $(0.7, 3.6)$    \\  
Type II & $(0.7, 1.0)$ & $(0.6,1.7)$   & $(0.7, 1.0)$    &$(0.7,1.3)$    &\cellcolor{gray!25} $(0.8, 3.2)$    &\cellcolor{gray!25} $(0.9, 58)$    \\  
Type X &  $(0.7, 1.0)$   & $(0.7,1.0)$  &  $(0.7, 1.1)$        &$(0.7, 1.1)$    &\cellcolor{gray!25} $(0.7, 21)$    & \cellcolor{gray!25} $(0.7, 23)$   \\
Type Z & $(0.7, 1.0)$  & $(0.6,1.7)$  & $(0.7, 1.0)$     &$(0.7, 1.1)$    &$(0.7, 1.1)$    & $(0.8, 1.2)$     \\
  \hline
\end{tabular}
\caption{ \sl Resulting intervals for the ratios obtained from the scans in various types of 2HDM.}
\label{tab:2} 
\end{table}

Based on the results shown in Tab.~\ref{tab:2}, one could conclude that the light CP-odd Higgs can modify the decay rates of the processes considered in this paper even in the case in which the muons (or electrons) are in the final state. However, this is mainly due to $\Gamma_h$ which in 2HDM can be larger than in the Standard Model due to $\Gamma(h\to \mathrm{invisible})\approx\Gamma(h\to AA)+\Gamma(h\to AZ)$. This width enhancement can be as large as $0.4 \times \Gamma_h^\mathrm{SM}$, which is why the ratios~\eqref{eq:RR} can be reduced by about $30\%$, cf. Tab.~\ref{tab:2}. The ratios of decay widths alone  
\bea
{\Gamma(h\to \eta_{cb}Z)^\mathrm{2HDM}\over \Gamma(h\to \eta_{cb}Z)^\mathrm{SM} } \quad \mathrm{and}\quad {\Gamma(h\to \eta_{cb}\mu^+\mu^-)^\mathrm{2HDM}\over \Gamma(h\to \eta_{cb}\mu^+\mu^-)^\mathrm{SM} },
\eea
remain either insensitive to 2HDM scenarios or only slightly enhanced, up to about $10\%$  with respect to their Standard Model values. In other words, the contribution arising from the diagram shown in Fig.~\ref{fig:fey3}c is indeed small.

\begin{figure}[htb]
  \includegraphics[width=0.35\textwidth]{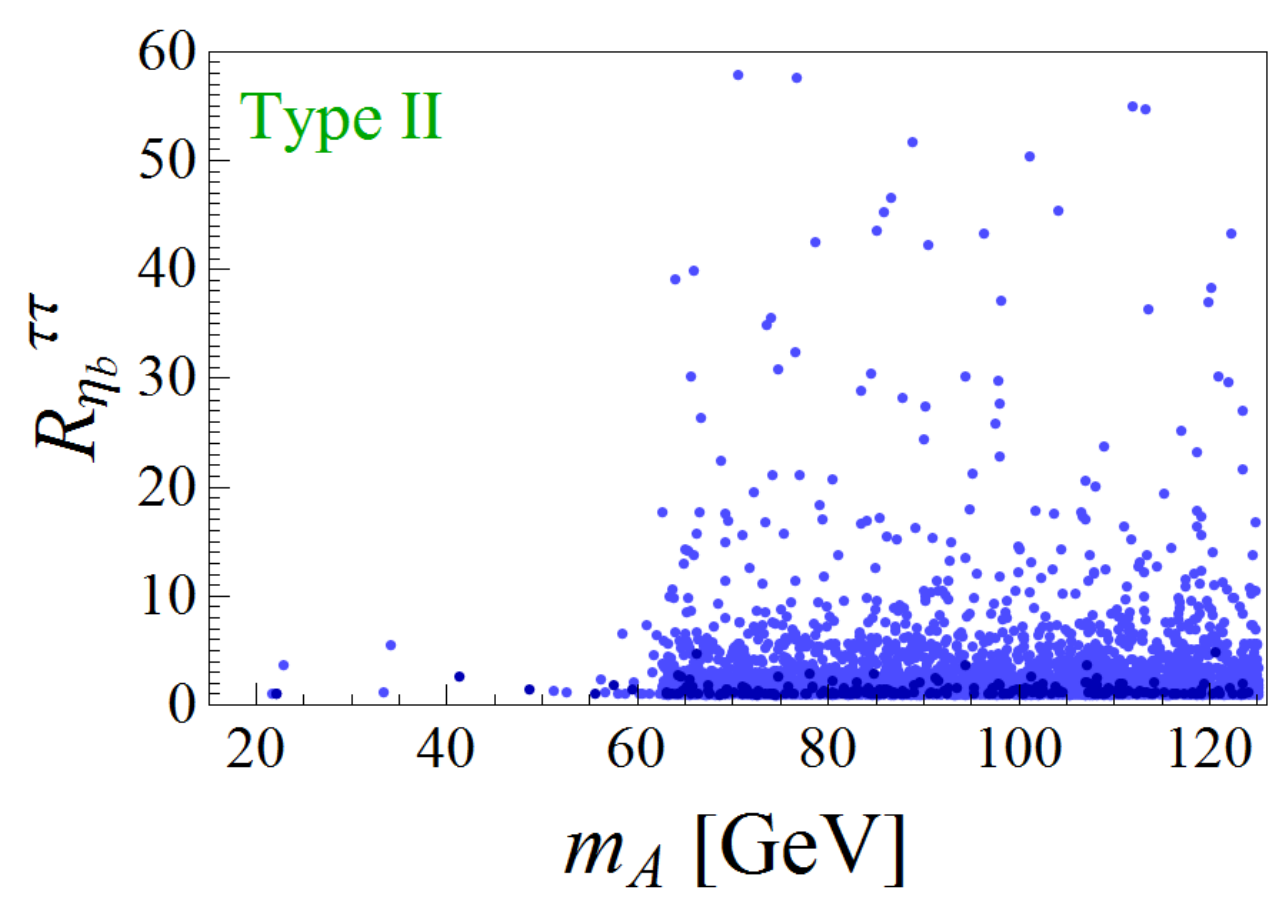}~\includegraphics[width=0.35\textwidth]{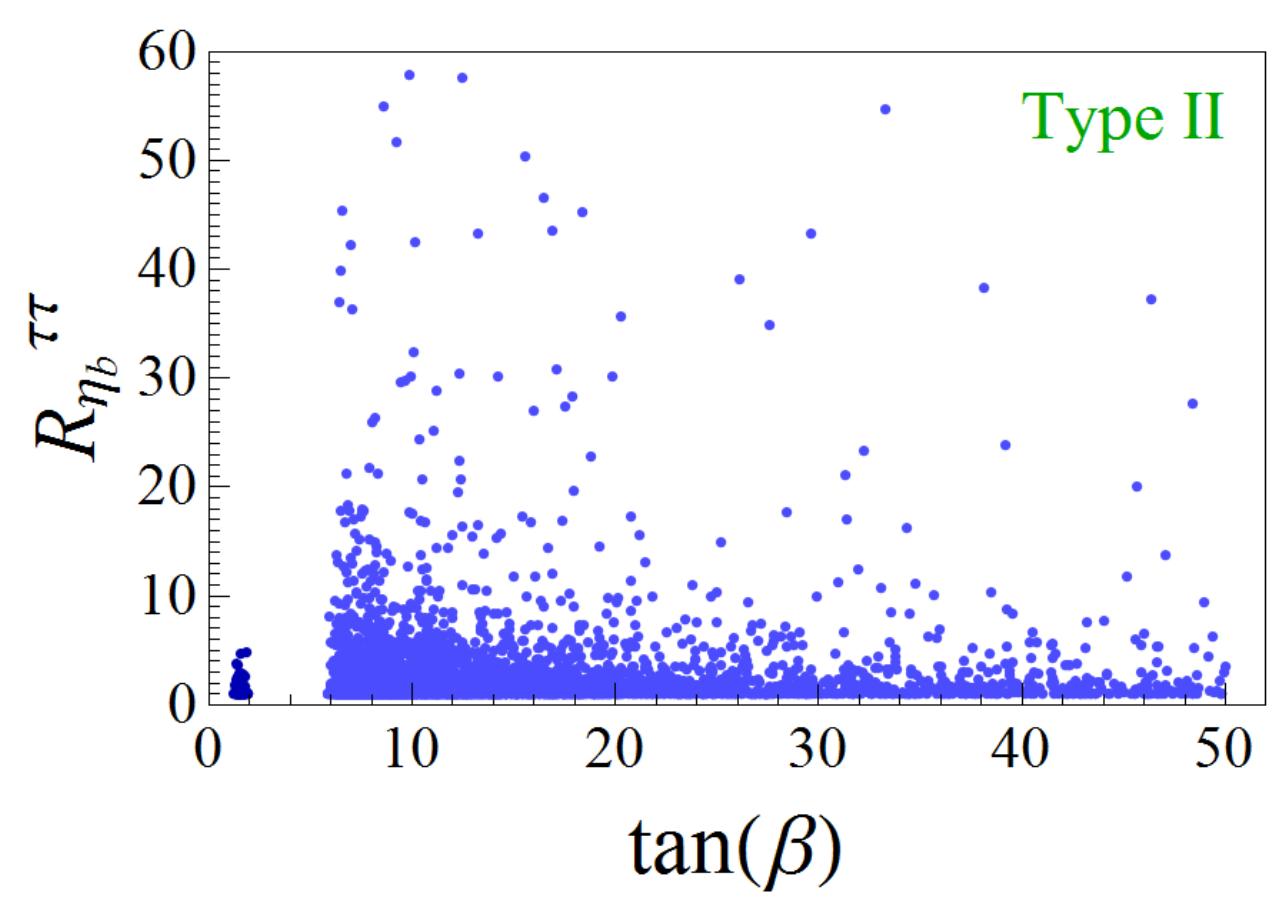}~\includegraphics[width=0.35\textwidth]{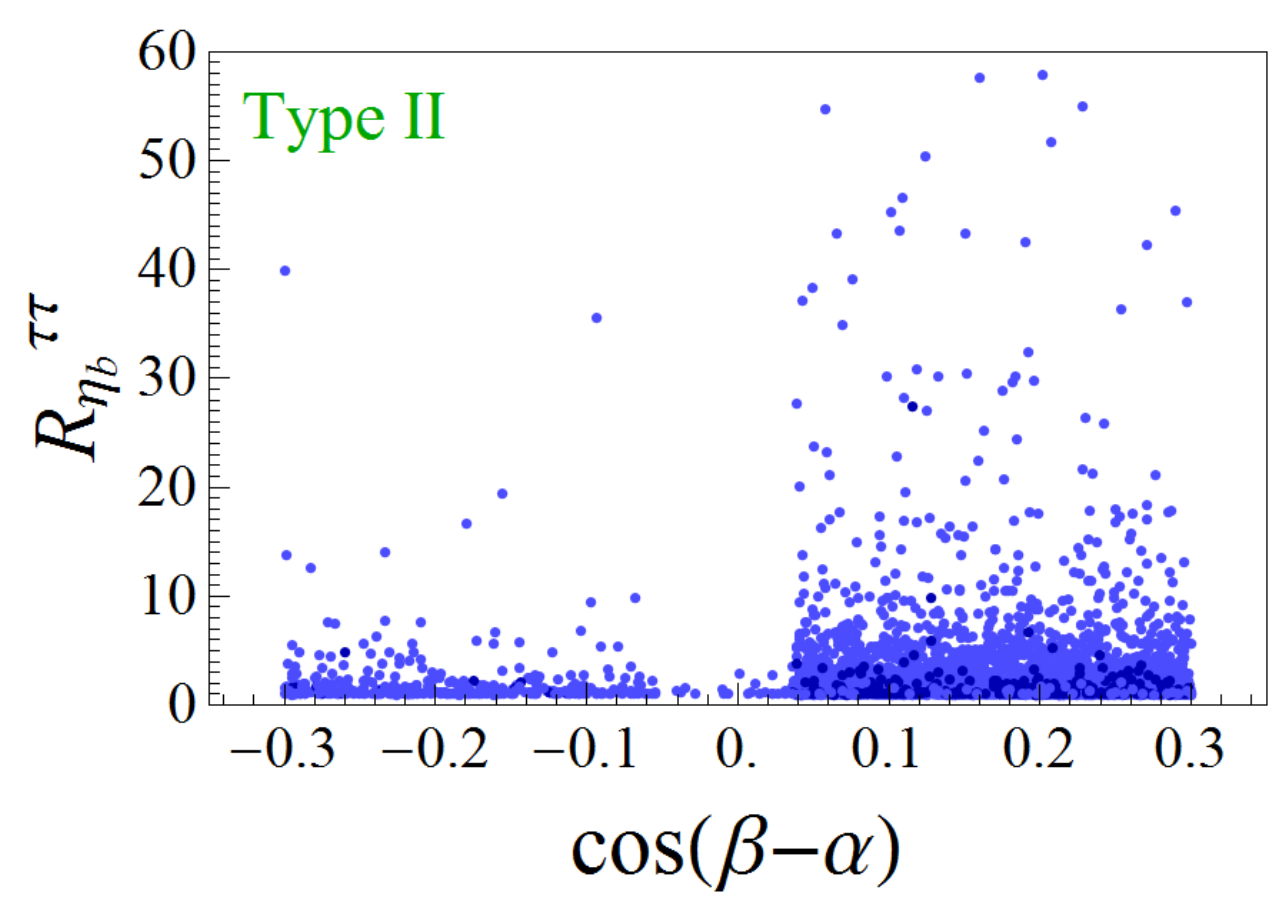}  \\
  \includegraphics[width=0.35\textwidth]{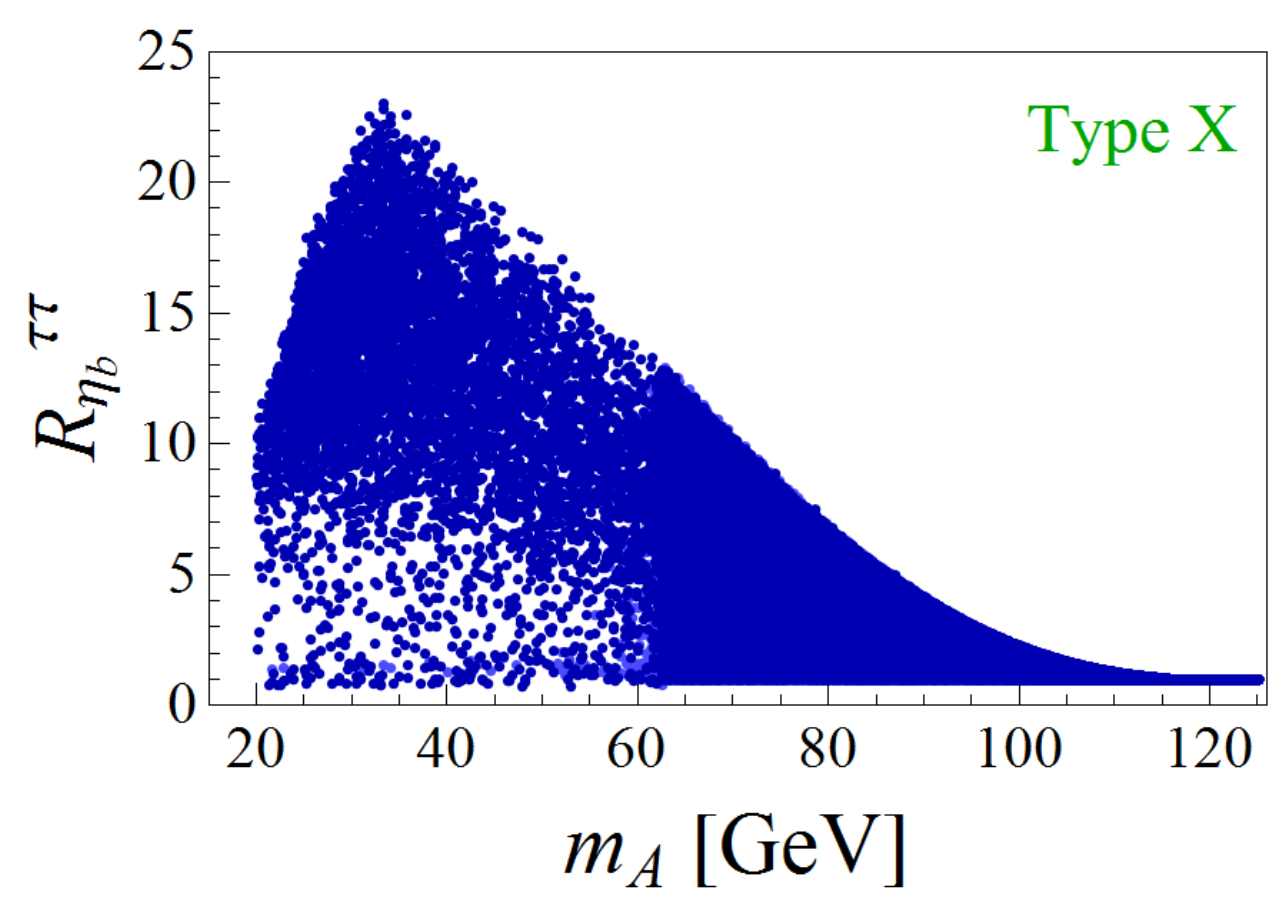}~\includegraphics[width=0.35\textwidth]{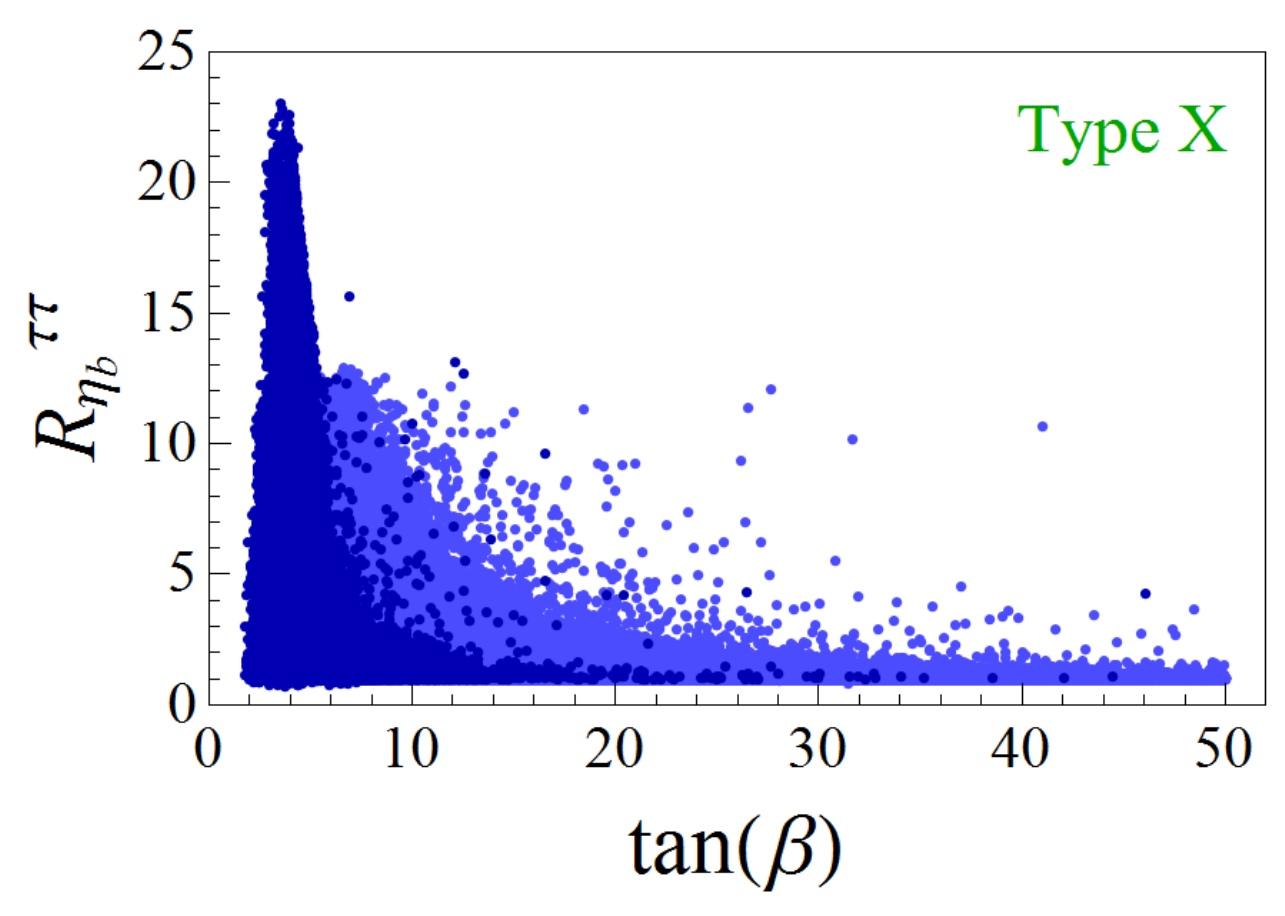}~\includegraphics[width=0.35\textwidth]{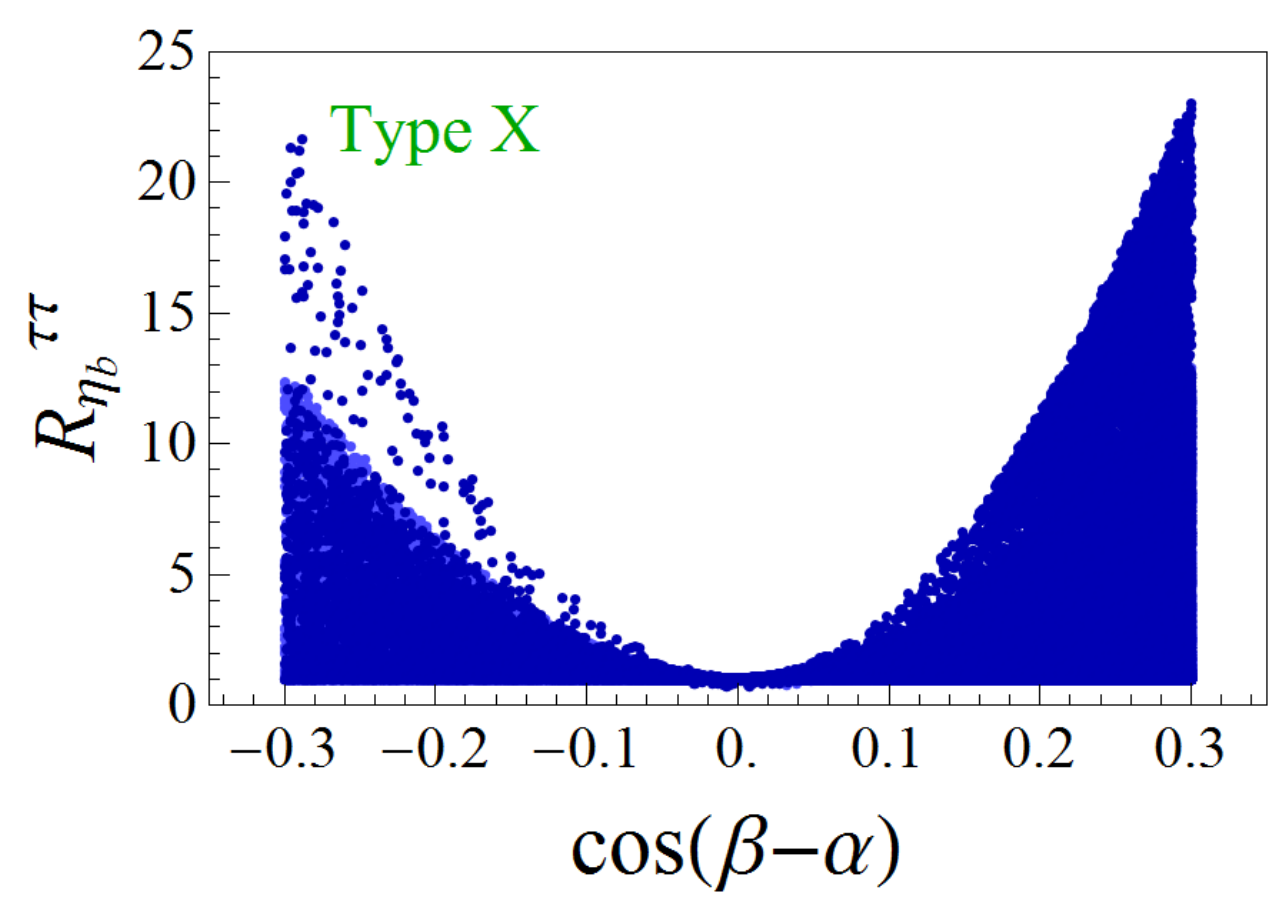}
  \caption{ \sl Results of the scan of parameters~\eqref{eq:scan2} after imposing constraints discussed in the text. Darker/lighter points correspond to the {\sl free}/{\sl fine-tuned} scan.  }
  \label{fig:6}
\end{figure}

In order to exacerbate the sensitivity to the CP-odd Higgs, one should consider $\tau$-leptons in the final state. This is because the second part in Eq.~\eqref{eq:total} becomes important, $\Gamma( h\to PA^\ast \to P \ell^+\ell^- )\propto m_\ell^2$, which can also be seen by using an approximate relation, $\Gamma (h\to \eta_{cb}\tau\tau) \approx \Gamma (h\to \eta_{cb} A)\ \cb (A\to  \tau\tau)$. Indeed, on the basis of the results presented in  Tab.~\ref{tab:2} we see that the ratios $R_{\eta_c}^{\tau\tau}$ and $R_{\eta_b}^{\tau\tau}$ depend much more on the light CP-odd Higgs than the ones with the light leptons in the final state. 
This is particularly true for the Type~I, II and~X models, the results highlighted in Tab.~\ref{tab:2}, and illustrated in Fig.~\ref{fig:6} for Type~II and X. We already stressed that the Type~II model is far more constrained than Type~X because of the constraint coming from $B\to X_s\gamma$. Yet the results for $R_{\eta_b}^{\tau\tau}$ exhibit the similar enhancement in both models, which can be traced back to $\Gamma( h\to PA^\ast \to P \ell^+\ell^- )\propto m_\ell^2\tan^2\beta$, a common feature of both models. Notice, however, that for larger values of $m_A$, the value of $\cb(h\to\eta_{cb}\tau^+\tau^-)$ rapidly approaches its Standard Model result, which is why we focus on these decay modes as possible probes of the light CP-odd Higgs ($m_A \lesssim m_h$).

%%%%%%%%%%%%%%%%%%%%%%%%%%%%%%%%%%%%%%%%%%%%%%
%%%%%%%%%%%%%%%%%%%%%%%%%%%%%%%%%%%%%%%%%%%%%%
%%%%%%%%%%%%%%%%%%%%%%%%%%%%%%%%%%%%%%%%%%%%%%
%%%%%%%%%%%%%%%%%%%%%%%%%%%%%%%%%%%%%%%%%%%%%%

\begin{figure}[htb]
  \includegraphics[width=0.5\textwidth]{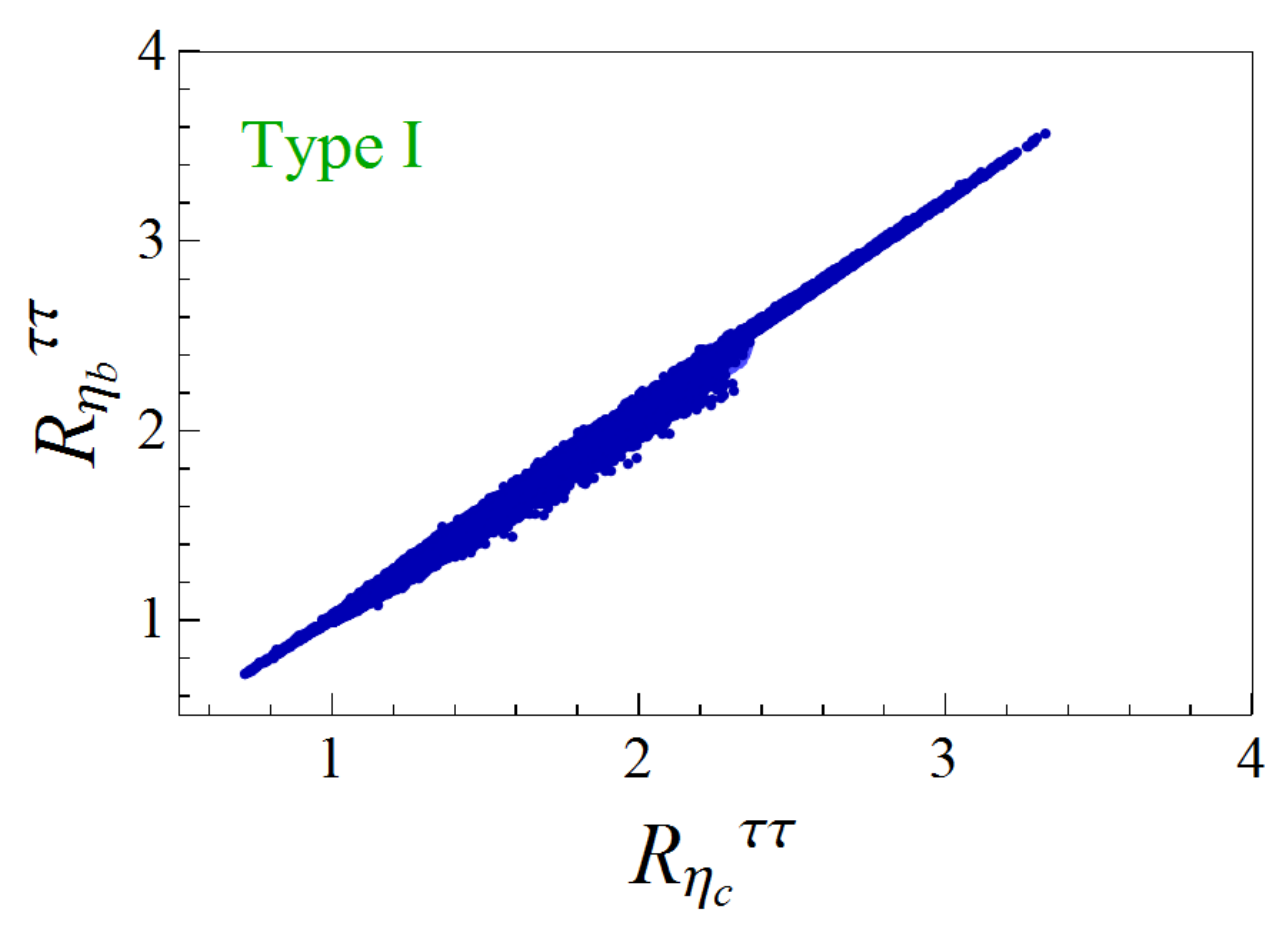}~\includegraphics[width=0.5\textwidth]{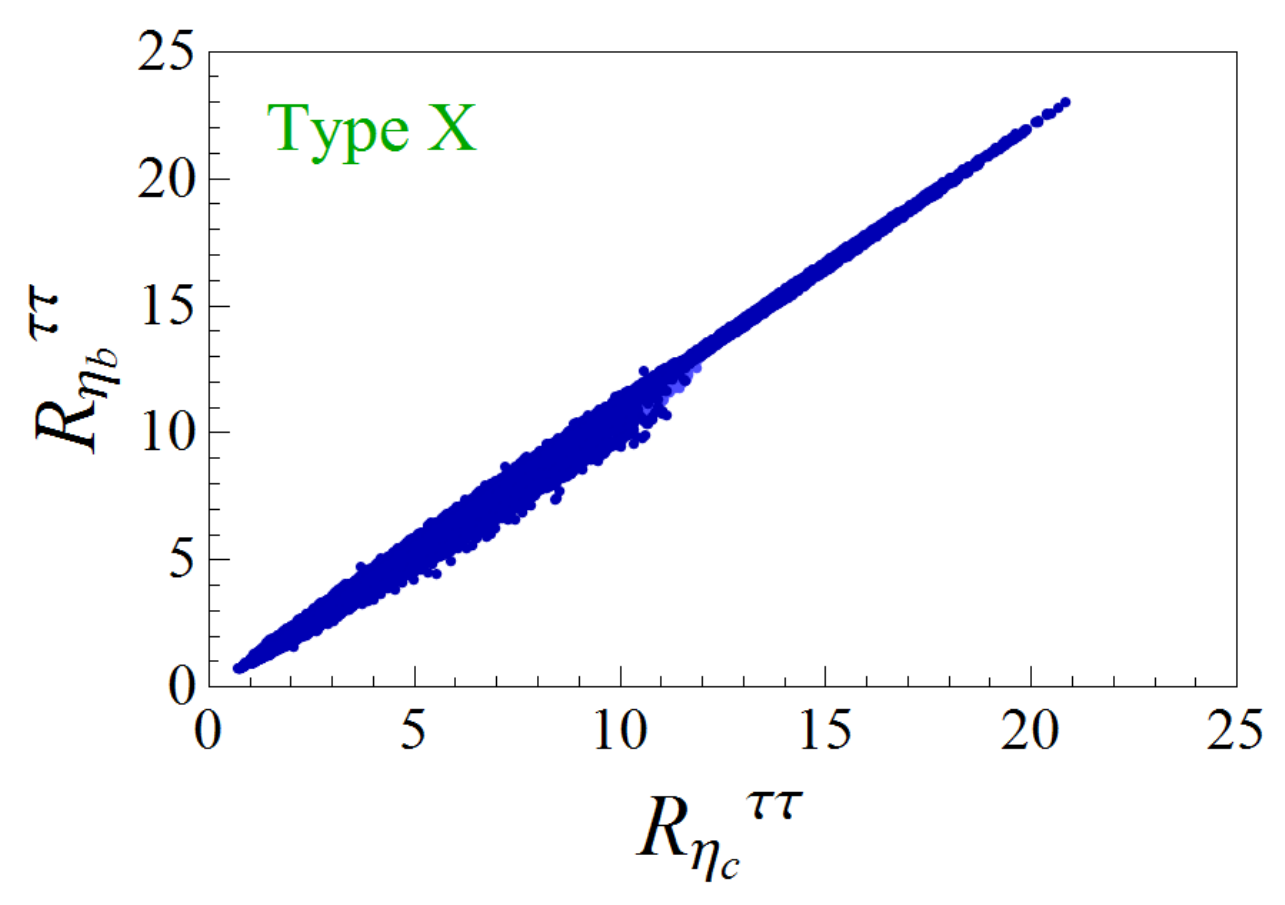}
  \caption{ \sl Correlation of the ratios $R^{\tau\tau}_{\eta_{c}}$ and $R^{\tau\tau}_{\eta_{b}}$ in Type~I and Type~X models arises from the fact that the Yukawa couplings of the charm and bottom quarks to the CP-odd Higgs are equal in these two models.  }
  \label{fig:7}
\end{figure}

Finally, it is worth mentioning a correlation between $R^{\tau\tau}_{\eta_{c}}$ and $R^{\tau\tau}_{\eta_{b}}$ in Type~I and Type~X models, which is shown in Fig.~\ref{fig:7}. It is easy to understand its origin once one realizes that $\Gamma( h\to PA^\ast \to P \ell^+\ell^- )$ dominates the full decay rate~\eqref{eq:total}, and since the couplings to charm and to bottom quarks are equal in both models, $|\xi^c_A|=|\xi^b_A|=1/\tan\beta$, the correlation becomes quite obvious. 
Similar reasoning, but this time with respect to $\xi^\tau_A$, can be used to explain why the enhancement in Type~X model ($|\xi^\tau_A|=\tan\beta$) is much more pronounced than the one in Type~I model ($|\xi^\tau_A|=1/\tan\beta$).
%%%%%%%%%%%%%%%%%%%%%%%%%%%%%%%%%%%%%%%%%%
%%%%%%%%%%%%%%%%%%%%%%%%%%%%%%%%%%%%%%%%%%%%%%
%%%%%%%%%%%%%%%%%%%%%%%%%%%%%%%%%%%%%%%%%%%%%%
%%%%%%%%%%%%%%%%%%%%%%%%%%%%%%%%%%%%%%%%%%%%%%
%%%%%%%%%%%%%%%%%%%%%%%%%%%%%%%%%%%%%%%%%%%%%%

\section{Summary and Conclusion}
\label{sec:concl}
In this paper we elaborated on a possibility to search the signal of a light CP-odd Higgs state through the decays 
$h\to PZ$ and $h\to P\ell^+\ell^-$, with $P$ being either $\eta_c$ or $\eta_b$.  
We derived the relevant expressions, identified the dominant parts of the decay amplitudes and then focused on various types of 
the 2HDM. In our scan of the 2HDM parameter space we fixed $m_h=125.09(24)$~GeV and varied other parameters in the ranges indicated in Eq.~\eqref{eq:scan2}, and then selected them as acceptable if consistent with a number of general theoretical constraints to which we also added
those coming from the exclusive $b\to s\mu^+\mu^-$ modes~\cite{Arnan:2017lxi} as well as the one arising from the inclusive $b\to s\gamma$ decay.  That latter mode provides a substantial shift of the lower bound on $m_{H^\pm}$ in Type~II and Type~Z models~\cite{Misiak:2017bgg}.

We find that most of the branching fractions are either consistent with their Standard Model values or can get lowered (increased) by about $30\%$ ($10\%$). Notable exceptions are the models of Type~I, II and X, in which the enhancement can be as large as a factor of $\sim 3$, and even a factor $\sim 20$, as highlighted in Tab.~\ref{tab:2} where we summarized the numerical results of our study. The origin of that enhancement is due to $m_A\lesssim m_h$ and it is related to the Yukawa couplings to the CP-odd Higgs, which explains why the effect is so pronounced in the case of $\tau$-leptons in the final state (in contrast to the case of muons or electrons).

The large enhancement of $\cb(h\to \eta_b\tau^+\tau^-)$ in Type~II and Type~X models as high as a factor $\approx 20$ with respect to its Standard Model value is also possible 
in the case of $\cb(h\to \eta_c\tau^+\tau^-)$, but only in the Type~X model, which is a consequence of the structure of Yukawa couplings and the amplitude mediated by $h\to AA$ vertex in the decay. 
While in this paper we focused on the decay of the SM Higgs boson, one could also consider a production of Higgs associated with the quarkonium state, along the lines similar to what has been discussed in Ref.~\cite{Brivio:2015fxa}.

We should also add that in this paper we focused on the lowest lying pseudoscalar quarkonia, but that our discussion could be trivially extended to the excited pseudoscalar quarkonia. Strategies for their detections have been discussed in Ref.~\cite{Godfrey:2015dia} and references therein.

\section*{Acknowledgments}
This project has received funding from the European Union's Horizon 2020 research and innovation program under 
the Marie Sklodowska-Curie grant agreements No.~690575 and No.~674896, as well as under the Twinning grant agreement No.~692194, RBI-T-WINNING. This work is also supported by the Croatian Science Foundation (HRZZ) project No.~5169, PhySMaB.

%\clearpage

\appendix

\section{Expressions for the amplitudes considered}
\label{app:angular}

We show here expressions for the direct contributions from diagrams in Figs.~\ref{fig:fey1}a,~\ref{fig:fey2}a and \ref{fig:fey3}a, and the indirect contributions involving leptons in Figs.~\ref{fig:fey1}b,~\ref{fig:fey2}b. 

The amplitude for the diagrams in Figs.~\ref{fig:fey1}a,~\ref{fig:fey2}a and   \ref{fig:fey3}a where the 
Higgs couples directly to the quark and anti-quark 
pair from which the meson is formed can be calculated 
in the QCD factorization approach, with the
bound state effect of the highly energetic hadrons 
in the final state accounted for in terms 
of light-cone distribution amplitudes (LCDA) of 
these hadrons. The light-cone projector for a 
pseudoscalar meson $P$ in momentum space, defined in terms of the matrix elements of the non-local quark 
and gluon current, up to twist-3 can be written as~\cite{Beneke:2001ev, Grossmann:2015lea},
\begin{eqnarray}
 M_P(k,x,\mu) &=& \frac{i f_P}{4}\left\lbrace\slashed{k} \gamma_5 \phi_P(x,\mu) -\mu_P(\mu)\gamma_5
 \left[\phi_p(x,\mu)- i \sigma_{\mu\nu} \frac{k^\mu \bar{n}^\nu}{k \cdot \bar{n}} 
 \frac{\phi'_\sigma (x,\mu)}{6} \right.\right. \nonumber \\
&& \left.\left. + i \sigma_{\mu\nu} k^\mu \frac{\phi_\sigma (x,\mu)}{6} \frac{\partial}{\partial k_{\perp \nu}}
\right] + 3{\rm -particle~ LCDAs}\right\rbrace ,
\end{eqnarray}
where $x$ is the longitudinal momentum fraction of meson carried by one of the valence quarks $q$, 
$\mu_P(\mu)=m_P^2/(2 m_q(\mu))$, $\phi_P$ is the twist-2 LCDA, $\phi_{p,\sigma}$ are the two-particle twist-3 LCDAs, while the three-particle ones are 
neglected which is often referred to as Wandzura-Wilczek approximation~\cite{Wandzura:1977qf}, which amounts to:
\begin{equation}\label{eq:WWA}
 \phi_p(x,\mu)  = 1,~~~~ \phi_\sigma(x,\mu)  = 6x(1-x).
\end{equation}
The light-like vector $\bar{n}$ in the above expression
is aligned into the opposite direction of $\vec{k}$.
%We choose the three components of $\bar{n}$ as the 
%$Z$ momenta ($\vec{p}_Z$).   {\red SIC!}

We have used  the following definition in the calculation of amplitudes for the diagrams in Figs.~\ref{fig:fey1}a,~\ref{fig:fey2}a and   \ref{fig:fey3}a which involve the integrals over the LCDA of the pseudoscalar meson:
\begin{align}\label{eq:prop}
I_1(x,y) &= \frac{1}{\mathcal{P}(x,1-x,y)}+ \frac{1}{\mathcal{P}(1-x,x,y)}, \quad
I_3(x,y) = \frac{1}{\mathcal{P}(x,1-x,y)^2}+ \frac{1}{\mathcal{P}(1-x,x,y)^2},\nonumber \\
I_2(x,y) &= \frac{x}{\mathcal{P}(x,1-x,y)}+ \frac{1-x}{\mathcal{P}(1-x,x,y)}, \quad
I_4(x,y) = \frac{x}{\mathcal{P}(x,1-x,y)^2}+ \frac{1-x}{\mathcal{P}(1-x,x,y)^2}, \nonumber \\
\bar{I}_2(x,y) &= \frac{1-x}{\mathcal{P}(x,1-x,y)}+ \frac{x}{\mathcal{P}(1-x,x,y)}, \quad
\bar{I}_4(x,y) = \frac{1-x}{\mathcal{P}(x,1-x,y)^2}+ \frac{x}{\mathcal{P}(1-x,x,y)^2},
\end{align}
where $\mathcal{P}(a,b,y) =  (a~m_h^2 + b~y^2 - a~b~m_P^2 - m_q^2)/m_h^2$. The expressions below are listed for the asymptotic form of the leading-twist LCDA, $\phi_P(x,\mu) = 6x(1-x)$.

The direct contribution to the $h\rightarrow P Z$ decay amplitude, including the light cone projector 
at leading and sub-leading twist is,
\begin{align}
 F^{PZ}_{D} &=  -f_P~\frac{m_q}{2m_h^2} 
 g_A^q  \int_0^1 dx \left[m_q \phi_P(x,\mu) I_1(x,m_Z) -\mu_P(\mu)\left\lbrace\phi_p(x,\mu) I_2(x,m_Z)
 \right.\right. \nonumber \\
 & \left.\left. +\frac{\phi_\sigma(x,\mu)}{6}
 \left(3 I_1(x,m_Z)-2 r_Z I_3(x,m_Z)-\frac{2}{m_h^2} (k\cdot p_Z) I_4(x,m_Z)\right)\right\rbrace \right]\nonumber \\
&= f_P~g_A^q~\frac{m_q}{m_h^2} 
 \left[2 \mu_P (\mu) - 3 m_q \right] \frac{1-r_Z^2+ 2 r_Z~{\rm ln}~r_Z}{(1-r_Z)^3}, 
\end{align}
where $g_A^q$ is the axial-vector coupling of the $Z$ boson to the constituent quark
of the meson, $k\cdot p_Z = m_h^2(1-r_Z-r_P)/2$ and $r_{X}=m_{X}^2/m_h^2$, $X=Z,P,q$. This direct contribution is completely negligible as it is suppressed relative to the leading term in Eq.~\eqref{eq:FPZ} by a factor of $m_P^2/m_h^2$ or $m_q^2/m_h^2$. 
We have also checked that the effect of the CP-violating coupling of the quarks to $h$ if present cannot be seen in these two body decays.

We next list the direct contribution to the three body Higgs decays Figs.~\ref{fig:fey2}a and~\ref{fig:fey3}a, along with the indirect contribution where the Higgs couples to the leptons  
Figs.~\ref{fig:fey2}b and \ref{fig:fey3}b. 
\begin{align}
\mathcal{M}(h\to P\ell^+\ell^-)^{2a} &= f_P \frac{m_q}{v m_h^2}
\left(\frac{g}{2 \cos \theta_W}\right)^2 \frac{g_A^q }{(q^2-m_Z^2+im_Z\Gamma_Z)}\nonumber \\
&\left[\left(\mathcal{C}_1~k^\mu  + \mathcal{C}_2~p^\mu \right)\left(-g_{\mu\nu}+\frac{q_\mu q_\nu}{m_Z^2}\right)
\left(\bar{u}_\ell \gamma^\nu (g_V^\ell-g_A^\ell \gamma_5)v_\ell   \right) \right], \label{eq:fig2a} \\
%%%%%%%%%%%%%%%%%%%%%%%%%%%%%%%%%%%%%%%%%%%%%%%%%%%
%%%%%%%%%%%%%%%%%%%%%%%%%%%%%%%%%%%%%%%%%%%%%%%%%%%
\mathcal{M}(h\to P\ell^+\ell^-)^{2b} &= f_P  \frac{m_\ell}{v}  \left(\frac{g}{2\cos\theta_W}\right)^2 
\frac{g^q_A}{(k^2-m_Z^2+im_Z\Gamma_Z)}
\left[k^\mu    \left(-g_{\mu\nu}+\frac{k_{\mu} k_{\nu}}{m_Z^2}\right) \right. \nonumber \\
& \left. \left\lbrace \frac{1}{(k+p_\ell)^2} 
\left(\bar{u}_\ell\gamma^\nu (g_V^\ell-g_A^\ell \gamma_5) (\slashed{k}+\slashed{p}_\ell)
v_{\ell}\right) 
\right.\right.\nonumber \\ &+ \left. \left.
 \frac{1}{(k+ p_{\bar{\ell}})^2} 
\left(\bar{u}_\ell (\slashed{k}+\slashed{p}_{\bar{\ell}}) 
\gamma^\nu (g_V^\ell-g_A^\ell \gamma_5)  v_\ell \right)\right\rbrace\right], \label{eq:fig2b} \\
%%%%%%%%%%%%%%%%%%%%%%%%%%%%%%%%%%%%%%%%%%%%%%%%%%
%%%%%%%%%%%%%%%%%%%%%%%%%%%%%%%%%%%%%%%%%%%%%%%%%%
\mathcal{M}(h\to P\ell^+\ell^-)^{3a}&=-f_P \frac{m_\ell m_q^2}{v^3m_h^2}\frac{\xi^\ell_A \xi^q_A }{(q^2-m_A^2+im_A\Gamma_A)} 
\mathcal{F}^{P\ell^+\ell^-}_A  \left(\bar{u}_\ell \gamma_5 v_\ell  \right),  \label{eq:fig3a}\\
%%%%%%%%%%%%%%%%%%%%%%%%%%%%%%%%%%%%%%%%%%%%%%%%%%
%%%%%%%%%%%%%%%%%%%%%%%%%%%%%%%%%%%%%%%%%%%%%%%%%%
\mathcal{M}(h\to P\ell^+\ell^-)^{3b}&=f_P\mu_P(\mu) \frac{m_\ell^2 m_q}{v^3}
\frac{\xi^\ell_A \xi^q_A }{(k^2-m_A^2+im_A\Gamma_A)} \nonumber \\
&\left\lbrace \frac{1}{(k+p_\ell)^2}
\left[\bar{u}_\ell\gamma_5 (\slashed{k}+\slashed{p}_\ell)
v_{\ell}\right]
%\right. & + \left.
+ \frac{1}{(k+p_{\bar{\ell}})^2}
\left[\bar{u}_\ell (\slashed{k}+\slashed{p}_{\bar{\ell}}) \gamma_5 v_\ell\right]\right\rbrace \label{eq:fig3b},
\end{align}
where we have
\begin{align}
\mathcal{C}_1 &=  f_P~\frac{m_q}{2m_h^2} 
 g_A^q \int_0^1 dx \left[m_q \phi_P(x,\mu)  I_1(x,q)
 -\mu_P(\mu)\left\lbrace -\phi_p(x,\mu) \bar{I}_2(x,q)
\right. \right. \nonumber \\
 & \left. \left.  
 +\frac{\phi_\sigma(x,\mu)}{6} \left(3 I_1(x,q) 
 -2 I_3(x,q)+2 \frac{(k\cdot p)}{m_h^2} \bar{I}_4(x,q) \right)\right\rbrace \right]  \\
 &= \frac{1}{m_h(1-r_q)^4}\left\lbrace(r_q-1) \left[6 m_q m_h (r_q^2-1) - \mu_P(\mu) 
\left(2 E_P (r_q+5) + m_h (r_q-1) (r_q+13)\right)\right] \right.\nonumber \\
&\left.
+2~{\rm ln}~r_q \left[2 E_P \mu_P(\mu) (2  r_q + 1) + m_h (r_q-1) (\mu_P(\mu) (4 r_q+3)-6 m_q r_q)\right]
\right\rbrace, \nonumber \\
 \mathcal{C}_2 &=  f_P~\frac{m_q}{2m_h^2}  g_A^q  \int_0^1 dx \left[
 -\mu_P(\mu)\left\lbrace\phi_p(x,\mu) I_1(x,q)
 \right.\right. \nonumber \\
 & \left.\left. \quad \qquad\qquad +\frac{\phi_\sigma(x,\mu)}{6}\left(2 \frac{(p_P\cdot p_h)}{m_h^2} 
  I_3(x,q)-2 r_P  \bar{I}_4(x,q) \right)\right\rbrace \right] \nonumber \\
  &=\frac{2\mu_P(\mu)}{m_h(1-r_q)^3}\left[2E_P(1+r_q)+m_h(1-r_q)^2+4E_P(1-r_q)\right]
  \,, \nonumber \\
\mathcal{F}^{P\ell^+\ell^-}_A &=\int_0^1 dx \left\lbrace \phi_P(x,\mu)
 \left[m_h^2 r_P I_2(x,q) + (k\cdot q) I_1(x,q)\right] -\mu_P(\mu) \phi_p(x,\mu) m_q I_1(x,q) \right\rbrace
 \nonumber \\
 &= -\frac{2}{(1-r_q)^3}\left[3\left((1-r_q^2)+2r_q {\rm ln}~r_q\right)(k\cdot q)+\mu_P(\mu)m_q 
 (1-r_q^2) {\rm ln}~r_q\right],
\end{align}
and $k\cdot q = 1/2 (m_h^2-m_P^2-q^2)$, $k\cdot p = m_h E_P$, with $E_P= 1/(2m_h) (m_h^2+m_P^2-q^2)$. The amplitudes for Fig.~\ref{fig:fey2}a [\ref{fig:fey2}b] listed in Eqs.~\eqref{eq:fig2a}~[\eqref{eq:fig2b}] being proportional to $m_q$ ($m_\ell$) can be safely neglected compared to Fig.~\ref{fig:fey2}(c), which is proportional to $m_Z$,  Eq.~\eqref{eq:fig2c}. 
The CP-violating coupling of the fermions to $h$ if present will contribute to $P\ell^+\ell^-$ final state, through the trilinear Higgs ($hhh$) coupling. We find that the $h\rightarrow P\ell^+\ell^-$ is insensitive to the CP-odd coupling, the reason being it enters quadratically and is proportional to the mass of the  final state fermions.

\end{document}